\documentclass{aastex63}

\received{\today}
\revised{\today}
\accepted{\today}
\submitjournal{ApJ}

\shorttitle{NGC 6240}
\shortauthors{Fyhrie et al.}

\begin{document}

\title{Molecular Gas in the Nuclear Region of NGC 6240}

\correspondingauthor{Adalyn Fyhrie}
\email{adalyn.fyhrie@colorado.edu}

\author{Adalyn Fyhrie}
\affil{CASA University of Colorado Boulder \\
	389 UCB \\
	Boulder, CO 80309, USA}

\author{Jason Glenn}
\affiliation{NASA Goddard Space Flight Center \\
	Code 665 \\
	8800 Greenbelt Road \\
	Greenbelt, MD 20771, USA}
\affiliation{CASA University of Colorado Boulder \\
	389 UCB \\
	Boulder, CO 80309, USA}

\author{Naseem Rangwala}
\affiliation{NASA Ames Research Center \\
	Space Science \& Astrobiology Division \\
	MS 245-6, Bldg 245, Rm 107F\\
	Moffet Field, CA 94035, USA}

\author{Jordan Wheeler}
\affiliation{CASA University of Colorado Boulder \\
	389 UCB \\
	Boulder, CO 80309, USA}

\author{Sara Beck}
\affiliation{Tel Aviv University\\
	School of Physics and Astronomy\\
	Ramat Aviv 69978, Israel}

\author{John Bally}
\affiliation{CASA University of Colorado Boulder \\
	389 UCB \\
	Boulder, CO 80309, USA}

\begin{abstract}
NGC 6240 is a luminous infrared galaxy in the local universe in the midst of a major merger. We analyze high-resolution interferometric observations of warm molecular gas using CO J = $3-2$ and $6-5$ in the central few kpc of NGC 6240 taken by the Atacama Large Millimeter Array. Using these CO line observations, we model the density distribution and kinematics of the molecular gas between the nuclei of the galaxies. Our models suggest that a disk model represents the data poorly.  Instead, we argue that the observations are consistent with a tidal bridge between the two nuclei. We also observe high velocity redshifted gas 
that is not captured by the model. These findings shed light on small-scale processes that can affect galaxy evolution and the corresponding star formation.
\end{abstract}



\section{Introduction} \label{sec:intro}

NGC 6240 (\cite{Wright1984}; \cite{Thronson1990}) is a unique galaxy in the local universe (\textit{z} = 0.02448) in the midst of a major merger event~\citep{Fried1983} that is triggering high star formation rates (\cite{Genzel1998};~\cite{Tecza2000}) and active galactic nuclei (AGN) activity in its two progenitor nuclei~\citep{Vignatti1999}, separated with a projected distance of around 1 kpc.
This activity results in a far-infrared (FIR) luminosity L$_{FIR} \approx $10$^{11.8}$ L$_{\odot}$ (\cite{Sanders1988}; \cite{Thronson1990};~\cite{Sanders1996}) that classifies it as a luminous infrared galaxy (LIRG), just below the threshold that would classify it as an ultra-luminous infrared galaxy (ULIRG). Its luminosity is expected to cross this threshold when a second starburst is triggered during final coalescence~\citep{Engel2010}. As such, NGC 6240 presents an excellent opportunity to study in fine detail the processes that power ULIRG activity.

The nuclear region ($\textless$1 kpc) of LIRGs and ULIRGs is a critical location to study as it is often the location of the highest star formation rates and the origin of many stellar feedback processes, such as stellar winds and AGN outflows. Correspondingly, in the most luminous LIRGs (L$_{IR} \textgreater $ 6$\times$10$^{11}$ L$_{\odot}$) the majority of the mid-IR emission originates from this central kpc region~\citep{Alonso2013}. 
The dynamics in this central region are complicated by the galaxy merger interactions that trigger the processes that lead to the high IR luminosity of most LIRGs~\citep{Lonsdale2006}. The nuclear molecular gas dynamics of NGC 6240 are no exception, with a concentration of molecular gas that peaks in emission \textit{between} the two nuclei while the stars and dust are found to be concentrated \textit{around} the two nuclei (\cite{Tacconi1999};~\cite{Tecza2000};~\cite{Engel2010}).

The molecular gas dynamics of NGC 6240's nuclear region have been studied and modeled for decades without consensus. \cite{Tacconi1999} observed a velocity gradient in CO J = $2-1$ in the nuclear region and modeled it as a rotating disk. A similar disk model was used to describe the motions of HCN by~\cite{Scoville2014}. This inter-nuclear disk model has both been used to support other authors' observations of molecular gas (e.g.~\cite{Iono2007} in CO J = $3-2$ and HCO$^+$ ($4-3$)) and been claimed to be unphysical in the context of other observations (e.g.~\cite{Gerssen2004} in H$\alpha$+[N II]). 
Alternate geometries have been proposed to explain this central molecular gas, including a tidal bridge connecting the two nuclei~\citep{Engel2010} and the origin site for a warm molecular outflow~\citep{Cicone2018}. Even more recently,~\cite{Treister2020} present observations of the nuclear region in CO J = $2-1$ in unprecedented detail with angular resolutions of 0.03$''$. They find the central region to be a clumpy concentration of gas dominated by a high velocity outflow and a gas bridge connecting the two nuclei and interacting with the stellar disk kinematics.

With the lack of consensus on the geometry of the inter-nuclear molecular gas, updated detailed modeling of the central nuclear region using multiple new line transition observations in high resolution is needed. New telescopes have allowed the nuclear region of NGC 6240 to be observed in unprecedented detail and have enabled this modeling. 
In this paper we use high resolution CO J = $6-5$ and CO J = $3-2$ observations from the Atacama Large Millimeter Array (ALMA) to model the velocity profile and density distribution of the central molecular gas using the Line Modeling Engine (LIME)~\citep{Brinch2010}. These higher-J transitions trace shock excited warm molecular gas that was detected but unresolved in observations by \textit{Herschel} (see, e.g.~\cite{Kamenetzky2014}). LIME produces emission line profiles based on three-dimensional velocity and density distributions provided to the modeling code. We alter these three-dimensional distributions to match the modeled to observed emission. We find multiple fiducial models, with the model whose emission most closely matches observations suggesting the central molecular gas is flat and pancake-like with a high velocity dispersion compared to its rotational velocity, suggesting that it is not a self-gravitating, rotating disk but instead a transient tidal bridge connecting the two nuclei. 

The residuals from the fiducial model definitively show high velocity gas ($\textgreater$ 300 km/s) that is not associated with the nuclear pancake-like concentration of gas. This high-velocity gas exists in others groups' observations and is potentially associated with outflows postulated therein (\cite{Cicone2018};~\cite{MullerSanchez2018};\cite{Treister2020}). The study of the high velocity gas is outside the scope of this work, which aims to describe the properties of the majority of the nuclear molecular gas. The presented observations also reveal extended emission features, some of which correspond to previously studied filamentary structures, and others of which are new to these observations. 

Section~\ref{sec:observations} presents the ALMA observations of CO J = $3-2$ and J = $6-5$ and continuum observations at 345 and 678 GHz. We analyze these observations in Section~\ref{sec:data analysis} and calculate the mass of the dust from the continuum emission, highlight observed extended molecular gas emission features, and explore the velocity structure of the gas. Section~\ref{sec: velocity structure} presents a non-local thermodynamic equilibrium model created using LIME that fits the nuclear molecular gas's velocity and density distributions. We find that the molecular gas between the two nuclei is unlikely to be a self-gravitating disk, but could instead be a tidal bridge in light of the model findings. 

\section{Observations}\label{sec:observations}

Observations of CO J = $3-2$ and J = $6-5$ were completed using ALMA with baselines of 1.6 km and 460 m, respectively. These are the first observations of the nuclear region of NGC 6240 in CO J = $6-5$ and the highest resolution observations to date in CO J = $3-2$. The CO J = $3-2$ observations were completed during Cycle 2 for project number 2013.1.00813.S. The CO J = $6-5$ observation was completed for project number 2015.1.00658.S during Cycle 3. The longest baselines, full-width half-maxima (FWHM) of the beams, channel widths,  reported channel root mean squared (RMS), central frequency of continuum observations, bandwidth (BW) of continuum observations, and RMS of continuum observations are reported in Table~\ref{table:observational parameters}. 
For the integrated moment 0 maps, $\sigma_{integrated}$ is calculated using line channel RMS values $\sigma_{line~channel}$ from Table~\ref{table:observational parameters}. The RMS in the moment map is calculated using $\sigma_{integrated}$ = $\sqrt{n_{chan}} \sigma_{line~channel}$ where $n_{chan}$ is the number of channels included in the integration.

To check the reliability of the provided reduced data products, we re-imaged the data using the Common Astronomy Software Applications (CASA) software package and the National Radio Astronomy Observatory (NRAO) provided imaging script. Re-imaging was completed with natural, uniform, and Briggs weighting schemes and user-created masks. For CO J = $3-2$ no new structure emerged in the molecular gas when re-imaging the data, nor could we improve upon the noise. Therefore, we deemed the NRAO provided data products sufficient for analyses of the CO J = $3-2$ molecular gas. 

Re-imaging was required for the CO J = $6-5$ and both continuum observations due to contamination in the continuum maps from extremely high-redshift with respect to systematic ($\sim$ 400 $-$ 740 km/s) molecular gas. This CO line contamination created a false continuum source between the two nuclei with the same peak flux density as the southern nucleus. Upon re-imaging, this source disappeared. We use Briggs weighting with robust parameter 0.5 for the re-imaged maps to match the weighting scheme used for the NRAO-provided molecular gas data products. 

Below, we introduce all observations with further details of all figures discussed in the following sections. At the distance of NGC 6240, 1$''$ corresponds to 500 pc of projected distance.  AGN locations from~\cite{Hagiwara2011} are included as crosses in the figures. 

\begin{figure}[ht!]
	\plotone{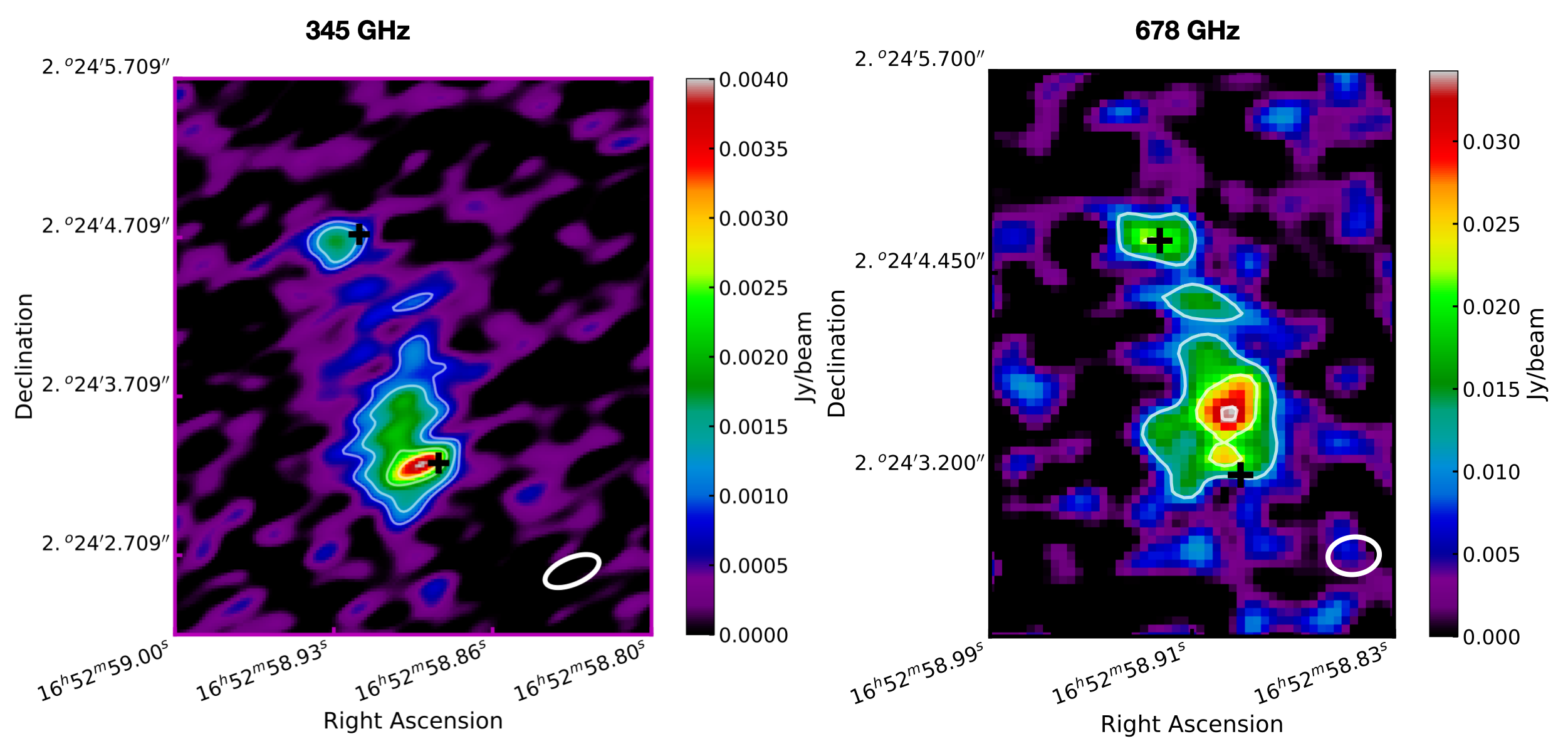}
	\caption{\textbf{Left:} \textit{Colors}: The 345 GHz continuum observation. \textit{Solid White Contours}: 4, 6, 10, and 14 $\sigma_{cont}$ contours. \textit{Black Crosses}: The locations of the two known AGN~\citep{Hagiwara2011}. The beam FWHM contour is plotted in the lower right of the image.  \textbf{Right:} \textit{Colors}: The continuum observation at 678 GHz. \textit{White Contours}: 5, 10, and 15 $\sigma_{cont}$ contours. \textit{Black Crosses}: The locations of the two known AGN. The beam FWHM contour is plotted in the lower right of the image, 1$''$ corresponds to 500 pc, east is left, and north is up.”}  \label{fig:continuum obs} 
\end{figure}

The two continuum maps are shown in Figure~\ref{fig:continuum obs}, revealing continuum emission in the vicinity of the nuclei as observed by previous authors (e.g. ~\cite{Scoville2014}). The concentration around the southern nucleus is much brighter than the northern concentration, with flux densities reported in Table~\ref{table:masses}. In detail, the morphologies appear to be different because of differing synthesized beams and signal-to-noise ratios in the two maps; however, there also differences that arise from real structure. In the south, the 678 GHz emission has a peak on the 345 GHz nucleus and in addition a stronger emission clump, $\sim$70$\%$ higher intensity, $\sim$0.4$''$ north of the southern nucleus and collinear with the two nuclei.  The bright clump was seen also by~\cite{Scoville2014}. This fidelity of continuum observations lends confidence to the associated line observations presented in this paper. The differences between the 678 and 345 GHz observations likely derives from a gradient of dust temperature and optical depth along the axis between the two nuclei. The 678 GHz source may arise in a  hot spot on the axis between the two nuclei or in a local fluctuation in the  optical depth; it should be observed further. The potential temperature and optical depth gradient, with temperature increasing to the north, is supported by the CO J = $6-5$ moment 0 centroid corresponding better than the CO J = $3-2$ moment zero centroid to the 678 GHz dust emission.

\begin{figure}[ht!]
	\plotone{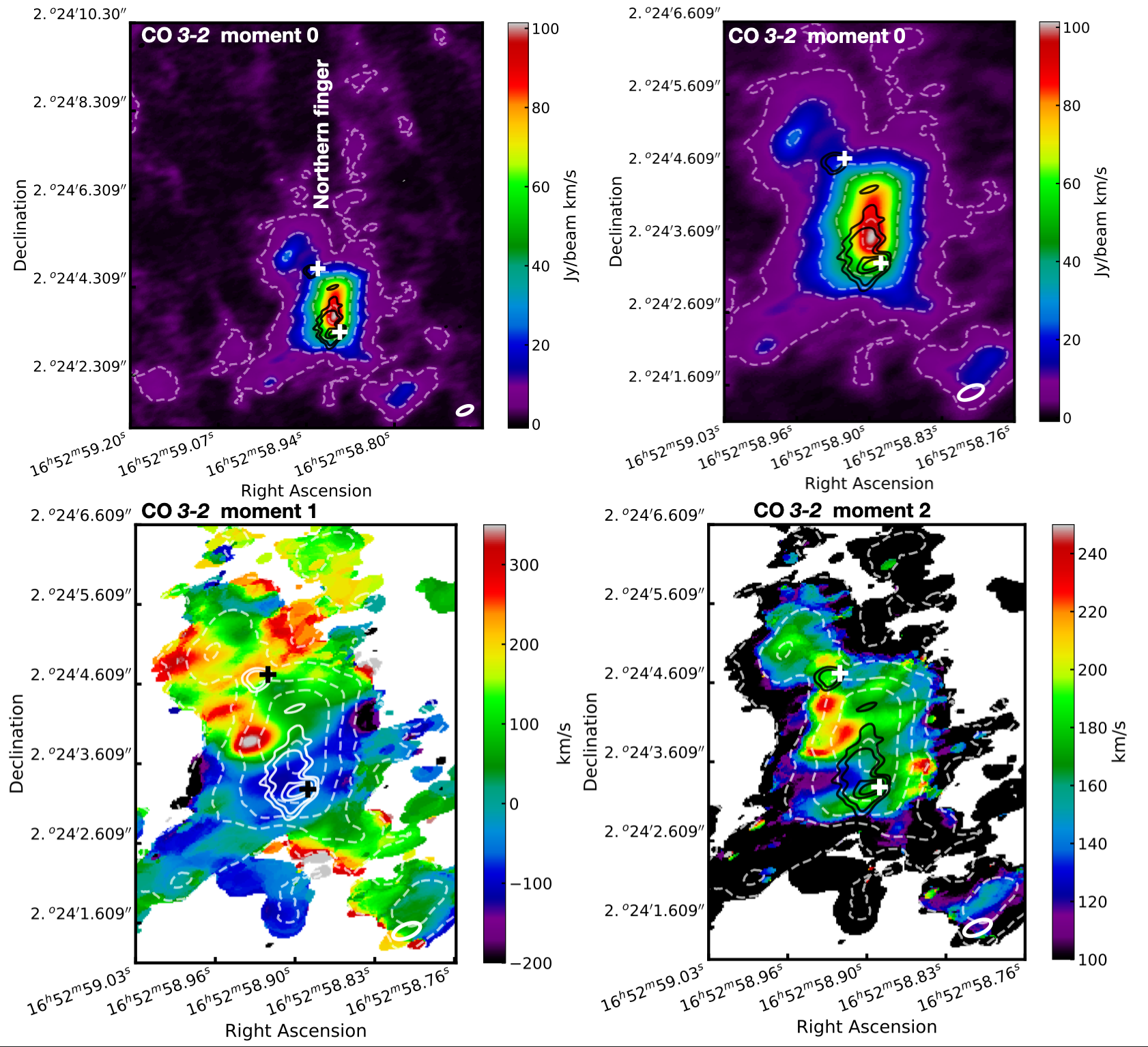}
	\caption{\textbf{Upper Left:} Full map of moment 0 of CO J = $3-2$ with \textit{dashed white contours} corresponding to 10, 25, 50, 100, 200, 300, 400, 500, and 600 $\sigma_{integrated}$ of the integrated line emission, \textit{solid black contours} show the 345 GHz continuum at 4, 6, 10, and 14 $\sigma_{cont}$, and \textit{crosses} show the location of the two AGN~\citep{Hagiwara2011}. \textbf{Upper Right:} The same, but zoomed to show only the nuclear region. \textbf{Lower Left:} Moment 1 in the nuclear region with channels below 5 $\sigma_{line~channel}$ masked out prior to moment calculation.  \textbf{Lower Right:} Moment 2 in the nuclear region with channels below 5 $\sigma_{line~channel}$ masked out prior to moment calculation. The beam FWHM contour is plotted in the lower right of each image, 1$''$ corresponds to 500 pc, east is left, and north is up. Velocities are calculated relative to the average radio velocity of CO J = $3-2$ for the entire observation, 7180 km/s, calculated relative to its rest frequency 345.796 GHz.\label{fig:CO32 moments}  }
\end{figure}

\begin{figure}[ht!]
	\plotone{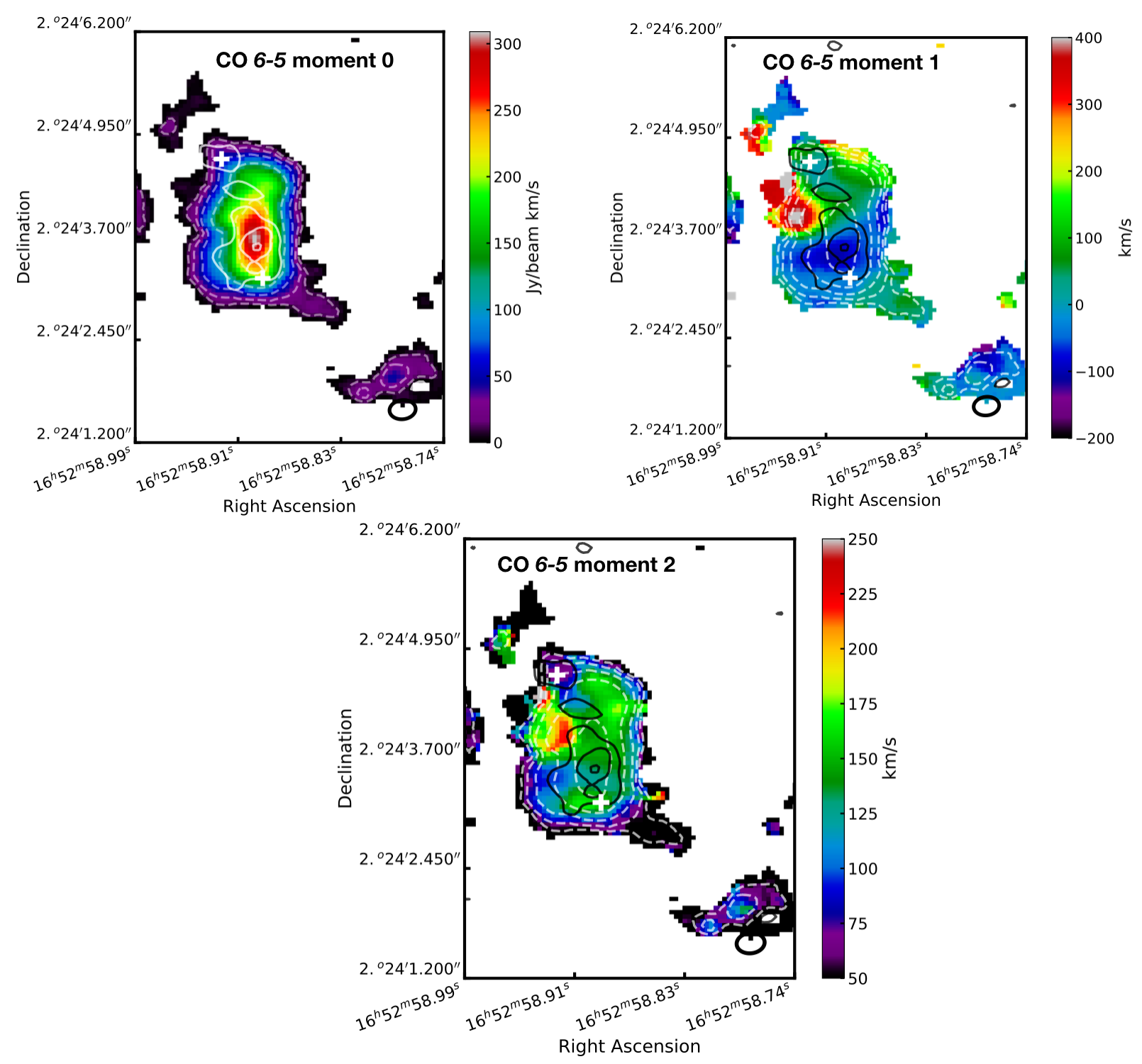}
	\caption{\textbf{Upper Left:} Moment 0 of CO J = $6-5$, with \textit{dashed contours} corresponding to 10, 25, 50, 100, 200 $\sigma_{integrated}$ of the integrated line emission and \textit{solid contours} show the 678 GHz continuum  at 5, 10, and 15 $\sigma_{cont}$. Channels below 5$\sigma$ are masked out of the integration. \textbf{Upper right:} Moment 1 of CO J = $6-5$, with pixels below 5 $\sigma_{line~channel}$ masked out prior to moment calculation. \textbf{Lower panel:} Moment 2 of CO J = $6-5$, with pixels below 5 $\sigma_{line~channel}$ masked out prior to moment calculation.  The beam FWHM contour is plotted in the lower right of each image. Channels below 5$\sigma_{line~channel}$ are masked out of the integration for all maps, 1$''$ corresponds to 500 pc, east is left, and north is up. Velocities are calculated relative to the average velocity of CO J = $6-5$ for the entire observation, 7150 km/s calculated relative to the CO J= $6-5$ rest frequency 691.473 GHz. \label{fig:CO65 moments}  }
\end{figure}

Moments 0, 1, and 2 (integrated flux density, average velocity, and velocity dispersion) for CO J = $3-2$ are plotted in Figure~\ref{fig:CO32 moments} with the 345 GHz continuum contours included. Similarly, Figure~\ref{fig:CO65 moments} shows the moments of CO J = $6-5$ with the 678 GHz continuum contours. Observations in both J = $3-2$ and J = $6-5$ show a concentration of molecular gas between the two nuclei, as observed previously in other observations of the molecular gas (e.g.~\cite{Tacconi1999},~\cite{Scoville2014},~\cite{Treister2020}). Also similar to previous observations, the CO moment 1 maps show a velocity gradient from highly redshifted with respect to systematic ($\textless v \textgreater \sim$ 350 to 400 km/s) to blueshifted with respect to systematic ($\textless v \textgreater \sim$ -100 to -150  km/s) along a position angle of approximately 34$^{\circ}$. Extended structure including two dim features of molecular gas to the SE and SW of the southern nucleus are observed in the CO J = $3-2$ maps that have also been observed in other wavelength bands, such as H$_2$~\citep{Max2005}, Fe XXV, and H$\alpha$~\citep{Wang2014} compared in Figure~\ref{fig:Max2005 comparison}.Although these lines trace different components of the ISM (molecular, shocked, ionized respectively) they are all associated with star formation activity. We also observe new extended emission to the north in the CO J = $3-2$ observation.

\begin{figure}[ht!]
	\plotone{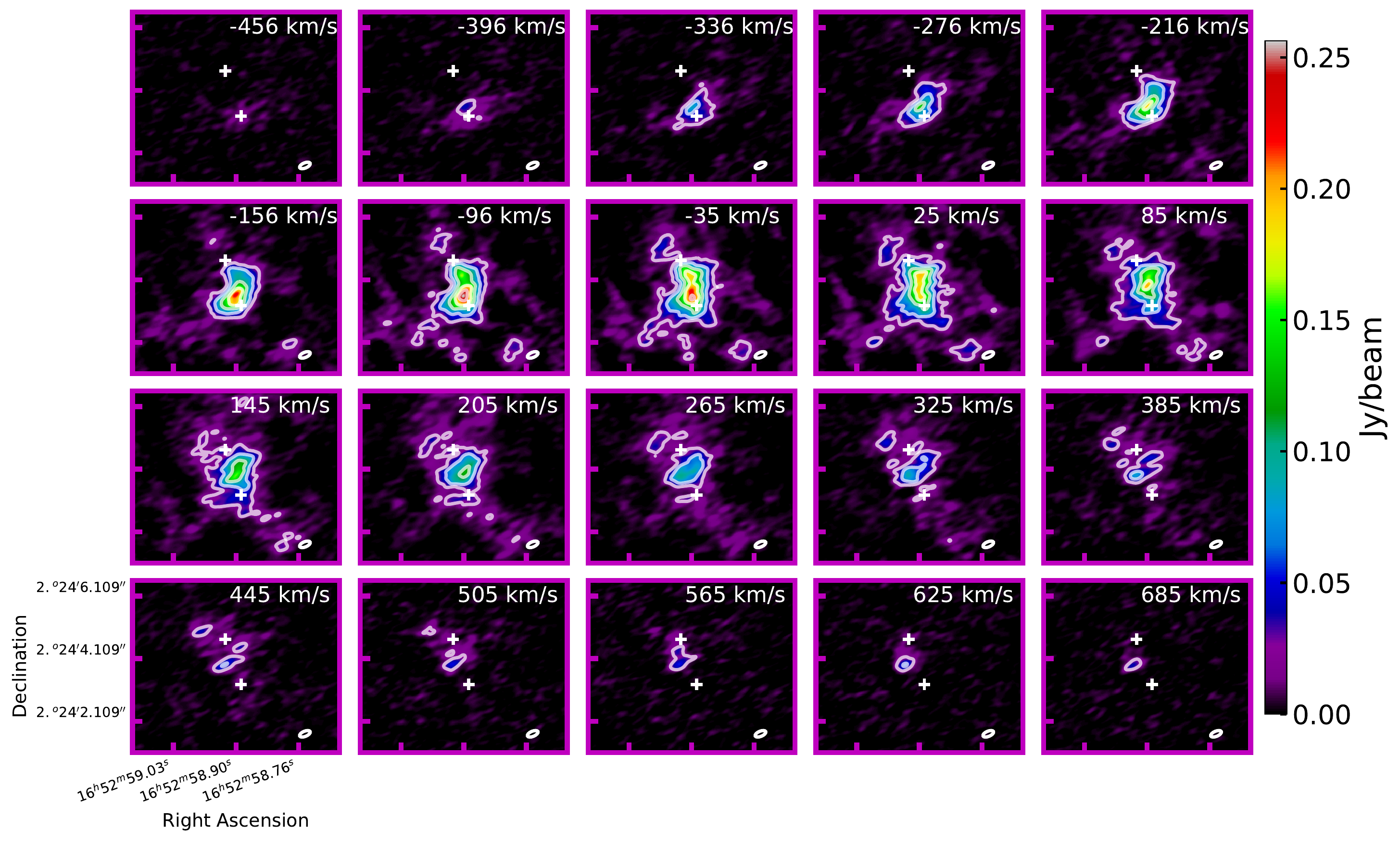}
	\caption{ The channel map of primary beam corrected, continuum subtracted CO J = $3-2$ observations show every third line channel (each panel is separated by 60 km/s). \textit{White contours:} 10, 25, 50, 75, and 100 times $\sigma_{line~channel}$ of the CO J = $3-2$ observations. \textit{White crosses:} the locations of the AGN. The beam FWHM contour is plotted in the bottom right of each figure, 1$''$ corresponds to 500 pc, east is left, and north is up. Velocities are calculated relative to the average velocity of CO J = $3-2$ for the entire observation, 7180 km/s calculated relative to its rest frequency 345.796 GHz. \label{fig: CO32_channelmap} }
\end{figure}

\begin{figure}[ht!]
	\plotone{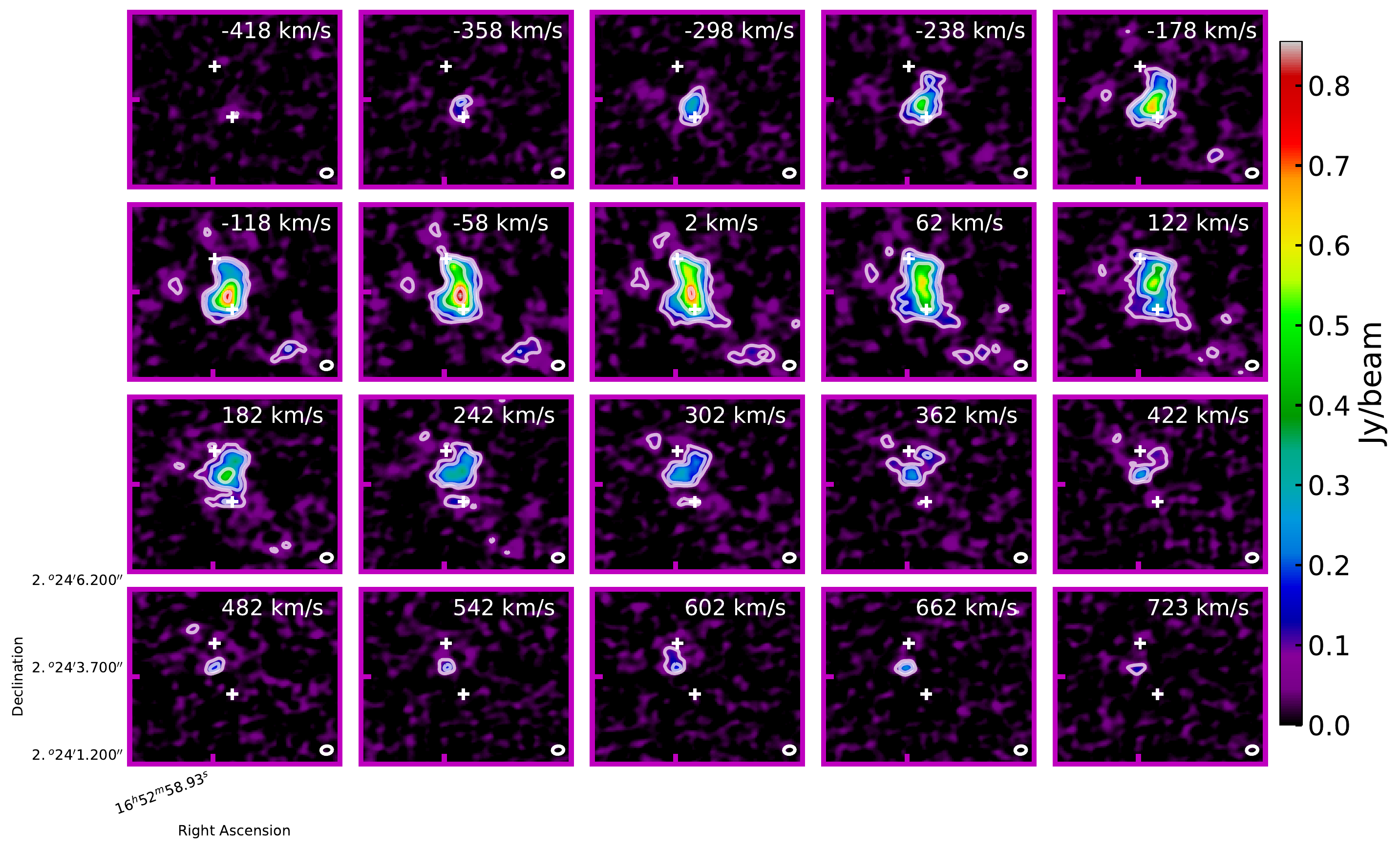}
	\caption{ The channel map of primary-beam-corrected, continuum-subtracted CO J = $6-5$ showing every third line channel. \textit{White contours:} 5, 10, 25, 50, and 75 times $\sigma_{line~channel}$ of the CO J = $6-5$ observations. \textit{White crosses:} The locations of the AGN. The beam FWHM contour is plotted in the bottom right of each figure, 1$''$ corresponds to 500 pc, east is left, and north is up. Velocities are calculated relative to the average velocity of CO J = $6-5$ for the entire observation, 7150 km/s calculated relative to its rest frequency 691.473 GHz. \label{fig: CO65_channelmap} }
\end{figure}

Channel maps for the CO J = $3-2$ and CO J = $6-5$ observations are shown in Figure~\ref{fig: CO32_channelmap} and Figure~\ref{fig: CO65_channelmap}, respectively. Radio velocities of the observations are calculated by the CASA software relative to the central observed frequencies, 345.796 GHz for CO J = $3-2$ and 692.2 GHz for CO J = $6-5$.   The systemic velocity of NGC 6240 is taken as the  velocity corresponding to the peak flux density of the galaxy-integrated line profile from the observations of CO J = $3-2$ and CO J = $6-5$: 7180 $\pm$ 20 km/s for CO J = $3-2$ and 7150 $\pm$ 20 km/s for CO J = $6-5$. 

The velocity structure of the inter-nuclear gas is most apparent in the higher sensitivity CO J = $3-2$ channel map, showing emission at  $\sim$-500 km/s with respect to systematic near the southern nucleus and emission at $\sim$ 650 km/s with respect to systematic located between the two nuclei. The CO J = $6-5$ channel map shows a similar velocity structure to that of the CO J = $3-2$.
This velocity structure is best resolved in the nuclear region between the two AGN, where the gas emission is brightest.

\startlongtable
\begin{deluxetable}{c|cc}
	\tablecaption{Parameters of observations received from ALMA. 
	\label{table:observational parameters}}

	\tablehead{ \colhead{Parameter} & \colhead{CO J = $3-2$ observation} & \colhead{CO J = $6-5$ observation} } 
	
	\startdata	
	Configuration (Baseline) & Extended (1.6 km) & Extended (460 m)\\ 
	Peak Molecular Flux Density Location & (16:52:58.898, +02.24.03.672) & (16:52:58.898, +02.24.03.756) \\
	Peak Flux Density & 100 Jy/beam km/s & 308 Jy/beam km/s\\
	Beam FWHM &  0.34 x 0.15$''$ & 0.32 x 0.23$''$ \\
	Beam PA & -67$^o$ & -83$^o$ \\
	Line channel width & 20 km/s & 20 km/s \\
	Line channel RMS $\sigma_{line~channel}$ & 2.52 mJy/beam & 14.5 mJy/beam \\
	Continuum band center $\nu$ & 345 GHz & 678 GHz \\
	Bandwidth & 7.5 GHz  & 6 GHz \\
	Continuum RMS $\sigma_{cont}$ &  0.21 mJy/beam &  2.1 mJy/beam \\
	\enddata
	
\end{deluxetable}

\section{Continuum and Extended Molecular Gas Emission Analysis}\label{sec:data analysis}

\subsection{Mass from Continuum}\label{sec:mass}
To calculate the mass of the dust $M_d$ from the continuum observations we use
\begin{equation}\label{eq:dust mass}
M_d = S_{\nu} D_L^2 / \kappa_{\nu} B_{\nu}(T) 
\end{equation}
\citep{Casey2012}, where $S_{\nu}$ is the flux density in the continuum frequency band, $D_L$ is the luminosity distance of 108 Mpc, $\kappa_{\nu}$ is the dust mass opacity coefficient, and $B_{\nu}(T)$ is the blackbody emission in the continuum frequency band for a dust temperature T. For this calculation, we use the 345 GHz as the measure of $S_{\nu}$ because it should have a lower optical depth than the 678 GHz continuum observation. The total $S_{345GHz}$ for this observation is 27 mJy, 18$\%$ of the galaxy-integrated flux density measured at 850 $\mu$m by SCUBA of 150 mJy~\citep{Klaas2001}. This flux density recovery is comparable to~\cite{Scoville2014} who measured 18-24 mJy at 340 GHz for a comparable beam size and sensitivity with ALMA in Cycle 0. The SCUBA beam is as large as our primary beam of the ALMA observations with a diameter of 15$''$~\citep{Klaas2001}, and as such we can expect that their measurement includes extended emission that is lost in our high-resolution observations. Therefore we expect our value of $S_{\nu}$ to be lower than that measured with SCUBA.

For $B_{\nu}(T)$ we choose a dust temperature of 56 K, the dust temperature fit in~\cite{Kamenetzky2014} using a greybody fit to \textit{Herschel}-SPIRE, IRAS, \textit{Planck}, SCUBA, and ISO photometry for NGC 6240. This dust temperature is a galaxy-averaged property since~\cite{Kamenetzky2014} used observations with beam FWHM between $\sim$ 17$''$ and 45$''$,  at least 100 times larger than the ALMA observations. It is likely that the dust temperature in NGC 6240's nuclear region is higher than the galaxy-averaged temperature due to the influences of concentrated star formation and AGN luminosity. However, we do not have flux density measurements across sufficient wavelengths in this central region to independently calculate the dust temperature. 
From~\cite{James2002} we use $\kappa_{850}$ = 0.07 m$^2$kg$^{-1}$ and $\kappa_{\nu} \propto \nu^2$ to find $\kappa_{870}$ = 0.067 m$^2$kg$^{-1}$. 

\begin{figure}[ht!]
	\plotone{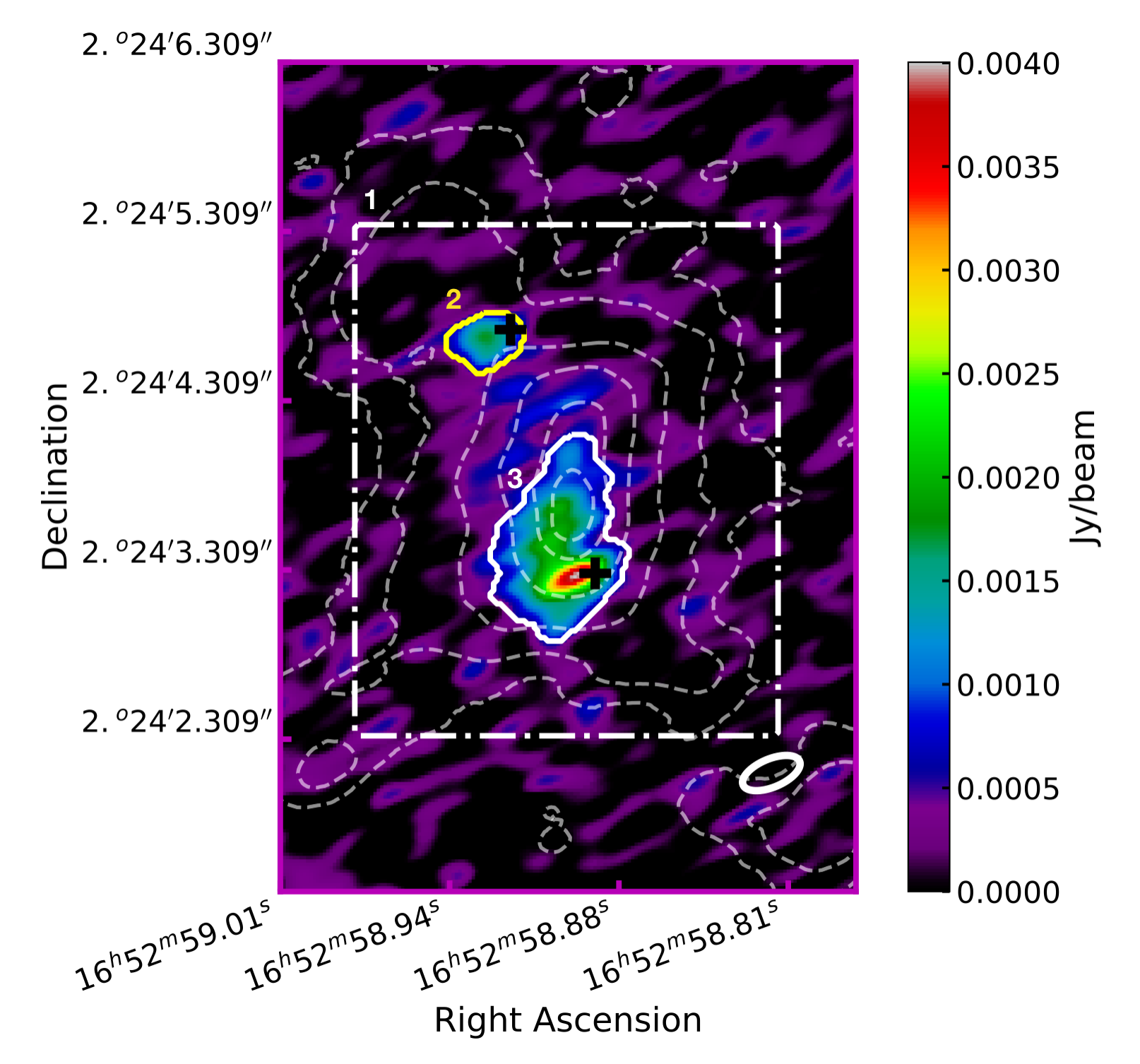}
	\caption{Regions for which we calculate the dust mass. \textit{Color map:} Extended configuration 345 GHz continuum map. \textit{Dashed contours}: CO J = $3-2$ integrated line emission. Masses for these regions are tabulated in Table~\ref{table:masses}. The beam FWHM contour for the 345 GHz continuum is plotted in the bottom right. \label{fig:mass regions} }
\end{figure}

The calculated masses for the regions outlined in Figure~\ref{fig:mass regions} are tabulated in Table~\ref{table:masses}.  The total mass of the dust is calculated to be 1.2$\times$10$^{7}$ M$_{\odot}$. As a check on this total dust mass we compare to the galaxy-integrated dust mass of 5$\times$10$^{7}$ M$_{\odot}$ found in~\cite{Kamenetzky2014} from their dust SED model described earlier in this section. Our calculated dust mass is 24$\%$ of this value, consistent with filtered flux from the ALMA observations.

Using a gas-to-dust mass ratio of 100 we can convert the dust masses from Table~\ref{table:masses} to a total gas mass, 1.2$\times$10$^9$ M$_{\odot}$. The mass values derived from the continuum emission are similar to the values derived from the sub-mm continuum (235 GHz) in~\cite{Treister2020} of 2.8$\times$10$^9$ M$_{\odot}$. The gas mass is converted to a column density $N_{H_2}$ by dividing the gas mass by the mass of molecular hydrogen, the projected size of the region on the sky, and multiplying by the fraction of the total mass assumed to be molecular Hydrogen (0.73, assuming solar metallicity~\cite{Hollenbach1987}). The average column density for the entire nuclear region is 3$\times$10$^{22}$ cm$^{-2}$, while the concentrations around the two nuclei both have higher column densities of 1.5$\times$10$^{23}$ cm$^{-2}$. This value is consistent with the findings of~\cite{Tacconi1999} of $N$(H$_2$) $\sim$ 1--2$\times$10$^{23}$ cm$^{-2}$.

\startlongtable
\begin{deluxetable}{ccccccc}
	\tablecaption{Integrated continuum flux densities $S_{\nu}$, derived dust masses from the 345 GHz continuum calculated using Equation~\ref{eq:dust mass}, gas masses assuming a gas-to-dust mass ratio of 100, and associated column densities of the regions outlined in Figure~\ref{fig:mass regions}. \label{table:masses}}
	\tablehead{
		\colhead{Region} & \colhead{Description} & \colhead{678 GHz $S_{\nu}$ [Jy]}  &\colhead{345 GHz $S_{\nu}$ [Jy]}  & \colhead{$M_d$ [M$_{\odot}$]} & \colhead{$M_{gas}$[M$_{\odot}$]} & \colhead{$N_{H_2}$ [cm$^{-2}$]}  }	
	\startdata
	1 &  Entire Continuum Observation & 0.27 & 0.027 & 1.2$\times$10$^{7}$ & 1.2$\times$10$^{9}$ & 4.0$\times$10$^{22}$\\
	2 &  Northern Concentration & 0.03 & 0.002 & 9.0$\times$10$^{5}$  & 9.0$\times$10$^{7}$ & 2.1$\times$10$^{23}$ \\
	3 & Southern Concentration & 0.116 & 0.013 & 5.8$\times$10$^{6}$ & 5.8$\times$10$^{8}$ & 2.4$\times$10$^{23}$ \\
	\enddata
\end{deluxetable}

\subsection{Extended Molecular Gas Emission}

A dim trail of gas extending directly north of the central concentration, which we call the ``Northern Finger", is observed in the higher-sensitivity CO J = $3-2$ observation but not in CO J = $6-5$ or either continuum observation. It is labeled in the CO J = $3-2$ moment 0 map in Figure~\ref{fig:CO32 moments}.  It is not spatially coincident with observed H$_2$ V = $1-0$ S(1) and S(5) emission (\cite{Max2005}, their Figure 12a). It dimly appears in their Figure 2a and 2c, false color images including Keck $K'$ band, $H$ band, $J$ band,  F814W filter from WFPC2 on $Hubble$ $Space$ $Telescope$ (HST), and the F450W filter from WFPC2 on $HST$. Its southern half is spatially coincident with [O III] $\lambda$ = 5007 and H$\alpha$ emission (\cite{MullerSanchez2018}, their Figure 1). 

The Finger's average velocity is 44 km/s  with a FWHM of $\sim$ 300 km/s, measured with an integrated spectrum of the emission in this region. The low average velocity suggests this gas is unlikely to be an outflow. A nuclear inflow would cause $\sigma_v$ to be elevated near the nuclei, for which there is no conclusive evidence. The low velocity is consistent with a tidal tail or bridge but the high FWHM does not conform to this idea. The finger extends from the northern nucleus towards the northern dust lane of the galaxy, which is interpreted as a tidal tail by other authors (\cite{Gerssen2004};~\cite{Yun2001}). The top edge of the finger is spatially coincident with the eastern edge of the dust lane, suggesting a possible correspondence. The dust lane is much wider and extends much farther beyond the extent of the finger -- approximately 7$''$ wide (roughly East/West) and over 20$''$ long (roughly North/South), beyond the extent of our observations.

\begin{figure}[ht!]
	\centering
	\resizebox{.5\linewidth}{!}{\plotone{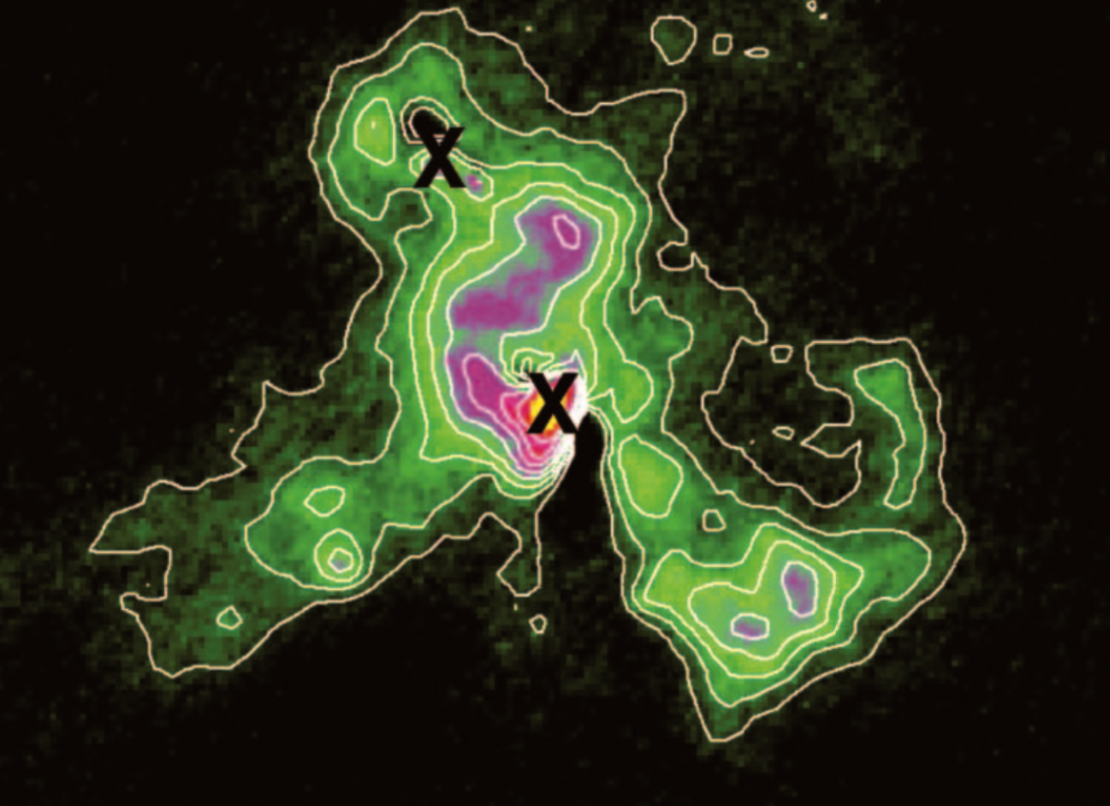}}
	\caption{ Figure 10 from~\cite{Max2005}, ``Keck AO narrowband image of the nuclear region in the H$_2$ 1-0 $S$(1) emission line [...] This image is 7.$''$5 wide by 5.$''$5 tall and has a log color map. The eight contour levels are on a log scale, with the first contour at 0.2 times the maximum flux (at the South1 nucleus) and the last contour at 0.02 times the maximum flux.'', included as a comparison to the CO J = $3-2$ moment 0 map in Figure\ref{fig:CO32 moments}. \label{fig:Max2005 comparison} }
\end{figure}

There is also extended structure surrounding the two nuclei that align with H$_2$ V = $1-0$ S(1) and S(5) concentrations presented in~\cite{Max2005} (their Figures 10 and 11, reproduced in our Figure~\ref{fig:Max2005 comparison}): a molecular concentration to the NE of the northern nucleus, a concentration to the SW of the southern nucleus, and dim, diffuse emission extending to the SE of the southern nucleus.  The faint arms of H$_2$ extending SE and SW observed in~\cite{Max2005} align with these faint arms extending SE and SW from the central concentration and are also bright in Fe XXV and H$\alpha$ \citep{Max2005, Wang2014}.~\cite{Wang2014} postulates that these faint molecular arms are molecular gas entrapped and shocked by the superwind caused by a vigorous starburst in the southern nucleus. Similarly,~\cite{Max2005} argue that these arms are a thin layer of gas at the edges of soft X-ray bubbles observed in~\cite{Komossa2003}, where a starburst driven wind is ``driving shocks or ionization fronts into the interstellar medium and surrounding molecular clouds". We therefore conclude that the CO arms to the SE and SW of the southern nucleus are likely associated with the same starburst-driven superwind shocked gas.

\section{Molecular Gas Between the Nuclei: a Test Model Motivated by History and Observations}\label{sec: velocity structure}

In~\cite{Tacconi1999} and in works since (e.g.~\cite{Iono2007};~\cite{Scoville2014}), the velocity gradient observed in the central molecular concentration was interpreted and modeled as a disk of gas between the two nuclei. The presence or absence of a molecular disk in the nuclear region of NGC 6240 is the basis for other authors' arguments regarding important stellar feedback processes, such as the outflow studied in~\cite{Cicone2018}. The disk model has been claimed to be unphysical in other studies of molecular gas in NGC 6240 (e.g.~\cite{Gerssen2004}), but no further modeling has definitively proven or disproven the internuclear disk model. 
More generally, the nuclear regions of (U)LIRGs are critical to understand as they are  often the location of the highest star formation rates and the origin of many feedback processes that affect star formation, such as stellar winds and AGN outflows~\citep{Alonso2013}.

For these reasons, we model this nuclear region as a concentration of gas with exponentially decreasing velocity and density profiles from the central peak using the observations in CO J = $3-2$ and J = $6-5$ to explore in detail the validity of this disk interpretation. The modeled gas concentration is not constrained to a self-gravitationally bound disk geometry, rather, it is a general concentration whose resultant velocity and density profiles can be compared to the disk interpretation. Our fiducial model finds the molecular gas concentration between the two nuclei is likely flat and pancake-like, with a high velocity dispersion that could cause it to dissipate as quickly as $\sim$ 0.3 Myr. Its geometry suggests this transient structure could be a tidal bridge connecting the two nuclei.

\begin{figure}[ht!]
	\plotone{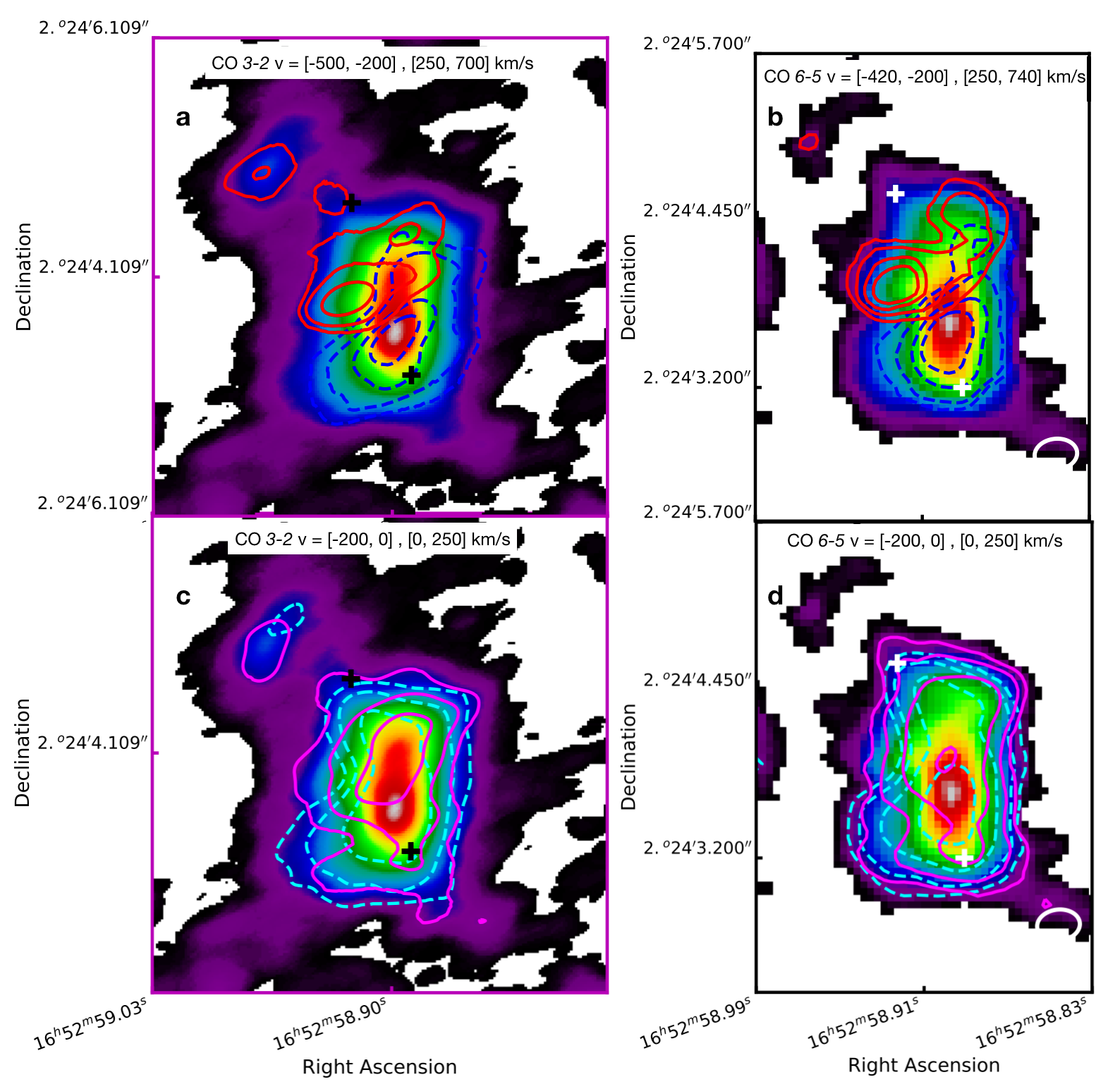}
	\caption{ Contours of high and moderate velocity redshifted and blueshifted gas for CO J = $3-2$ and J = $6-5$, plotted over corresponding integrated line emission (moment 0). Contours are 0.2, 0.4, 0.6, and 0.8 of the maximum in each map. Beam FWHM contours are plotted in the bottom right of panels b and d. \textit{Crosses} denote locations of the AGN.   \textbf{a}: CO J = $3-2$ high velocity gas; \textit{red contours:} [250, 700]  km/s, \textit{dashed blue contours:} [-500,-200] km/s   \textbf{b:}  CO J = $6-5$ high velocity gas; \textit{red contours:} [250, 740]  km/s, \textit{dashed blue contours:} [-420,-200] km/s   \textbf{c:} CO J = $3-2$ moderate velocity gas; \textit{red contours:} [0, 250]  km/s, \textit{dashed blue contours:} [-200,0] km/s  \textbf{d:} CO J = $6-5$ moderate velocity gas; \textit{red contours:} [0, 250]  km/s, \textit{dashed blue contours:} [-200,0]. \label{fig:redblueshifted} }
\end{figure}

\subsection{Code for Generating the Test Model: the Line Modeling Engine (LIME)}
To generate the molecular gas model that tests the validity of the disk interpretation, we use the Line Modeling Engine (LIME)~\citep{Brinch2010}. LIME uses non-local thermodynamic equilibrium radiative transfer and molecule rotational energy level populations calculations to predict the line and continuum emission from molecular clouds. The user defines a 3-D model describing the density distribution of hydrogen molecules of the source, then assigns 3-D temperature and velocity distributions to those molecules. The gas-to-dust mass ratio, the abundance of CO relative to hydrogen, and the 3-D distribution of the dust temperature are also set by the user. Finally, the user defines the distance to the source to obtain the proper distance scale per pixel and observed flux density. 

The code does not require that the source is in local thermodynamic equilibrium, and instead solves for population levels iteratively until the model populations have converged at all grid points. After convergence, LIME ray-traces photons to obtain an image of the modeled source at a user-defined observing angle. This simulation methodology allows great flexibility in both geometry and kinematics of the simulated source and minimizes the assumptions made about the source to generate the observed line profiles. 

To compare the observed data to the simulated model images, we use the CASA package to smooth and continuum subtract the simulated images. The resulting images have the same beam size as the observations. We must also account for the redshift of the galaxy, which is not included in the LIME simulations. To do this, we measure the central velocity of the galaxy-integrated line profile from the observations of CO J = $3-2$ and CO J = $6-5$: 7180 $\pm$ 20 km/s for CO J = $3-2$ and 7150 $\pm$ 20 km/s for CO J = $6-5$. These velocities are subtracted from the observed moment 1 maps in Figures~\ref{fig:CO32 moments} and~\ref{fig:CO65 moments}, which we can compare to the simulations that are simulated with zero redshift.

\subsection{Parameters of the Test Model}\label{sec:model parameters}
Using LIME, we model the 3-D central gas concentration with density and velocity distributions 
with exponentially decaying density and velocity profiles. Our 3-D model for the disk density follows the function
\begin{equation}\label{eq:nH2}
n_{H2} = n_{H_{2},0}  \exp(-r/r_{sh})  \exp(-|z|/z_{sh}) ,
\end{equation}
where $r = \sqrt{x^2+y^2}$, $r_{sh}$ is the radial scale height of the density, $z_{sh}$ is the vertical scale height, and $n_{H_{2},0}$ is 
the central
H$_2$ number density 
of the gas concentration. The total gas mass within the extent of our ALMA observations is set to a simulation parameter $g_{mass}$, from which $n_{H_{2},0}$ is calculated based on the simulation geometry.
We also allow an overall density asymmetry by multiplying the density on one half of the gas concentration (x$\textgreater$0) by a model parameter $a_{n}$. For the CO J = $3-2$ model we also allow the extent of the gas to differ for x$\textgreater$0 and x$\textless$0 by varying the scale height on either side of the modeled concentration.

We model the velocity with
\begin{equation}\label{eq:v}
v(r) = v_{circ} (1 - \exp(-r/r_{circ})) ,
\end{equation}
where $v_{circ}$ is the circular velocity and $r_{circ}$ is the radial scale length of the circular velocity. We use the exponential profile because sside from the high velocity gas component, we find no sufficient justification for deviation from a rotation curve that rapidly becomes flat with radius, (which would be appropriate in the case of a self-gravitating disk).

In Equations~\ref{eq:nH2} and~\ref{eq:v}, parameters except $r$ and $z$ are fit parameters that can be tuned to match the resultant modeled line profiles to the observed data. In addition to those parameters, the LIME model requires a gas inclination, position angle, dust temperature $T_{dust}$, turbulent velocity $v_{turb}$, gas temperature $T$, and gas-to-dust mass ratio (set to 100). All available parameters, fiducial fit results, and acceptable ranges of fiducial fit parameters can be found in Table~\ref{table:disk values} and are discussed in Section~\ref{sec:best fits}.

The position angle of the gas concentration is modeled as 34$^{\circ}$,
the angle between the maximum of the highly blueshifted gas and that of the highly redshifted gas from Figure~\ref{fig:redblueshifted}. This is close to the position angle used by~\cite{Tacconi1999} of 40$^{\circ}$ for their disk models. 
Thirty-four degrees was chosen following methodologies laid out in~\cite{Cicone2018} and~\cite{Tacconi1999}. Following the methodology presented in~\cite{Cicone2018}, we separate the observations into ``quiescent'' [-200, 250] km/s gas and high velocity gas. Figure~\ref{fig:redblueshifted} shows the contours of the observed highly redshifted (CO J = $3-2$: [250, 700] km/s,  CO J = $6-5$: [250, 740] km/s) and highly blueshifted (CO J = $3-2$: [-500, -200] km/s,  CO J = $6-5$: [-420, -200] km/s) gas in panels a and b. The centroids between these two velocity extremes (34$^{\circ}$) is at a steeper angle than between the two nuclei ($\sim$19$^{\circ}$).
The contours of moderate velocity, or ``quiescent'' gases of velocity [-200, 250] km/s are shown in in panels c and d. The velocity gradient observed in moment 1 for CO J = $3-2$ and J = $6-5$ remains clearly present in the high velocity gas, with distinct centroids of emission at a PA of $\approx$ 34$^{\circ}$. The lower velocity gas has much more overlap between the redshifted and blueshifted components, without as clear of a separation between centroids of emission. This lower velocity gas aligns approximately with the semi-major axis of the central gas concentration at a PA of $\approx$ 0$^{\circ}$, along the north/south axis. Due to the distinct gradient present in the high velocity gas that more closely resembles a traditional disk, we choose the position angle based on the highest velocity gas. However, we find during modeling that the position angle does not influence the fit parameters outside of acceptable ranges already incorporated into the model.

The center of the modeled gas concentration is placed at the maximum of the observed integrated line emission, following the methodology of~\cite{Tacconi1999}. If the gas were a simple disk, the maximum of the line emission would be expected to align closely with the turnover point of the velocity gradient. As noted in~\cite{Tacconi1999} and confirmed in this paper's observations, this is not the case. This offset could be caused by a number of phenomena, including that the gas is not a simple disk. The offset could also be explained by a gas concentration with a highly asymmetric mass distribution leading to an offset in the maximum line emission. Optical depth effects could also displace the velocity gradient turnover point from the maximum of the line emission.

\subsection{Test Model Fitting Strategy and Fiducial Fits}\label{sec:best fits}

In NGC 6240 the CO J = $6-5$ emission has been found to be dominated by relatively hot gas while the CO J = $3-2$ emission is dominated by by cool/warm gas~\citep{Kamenetzky2014}. The upper levels of the CO J = $6-5$ and J = $3-2$ CO lines are at very different energies, with the J = $6$ level at $\sim$116 K and the J = $3$ level at $\sim$33 K; it is clear that J = $6-5$ line will arise in warmer gas than does the J = $3-2$. The true excitation picture is of course more complicated than a simple model of `warm' and `cool' components. The J = $3-2$ line is to some extent sensitive to warm and hot gas: it is observed in quite hot regions (e.g.~\cite{Consiglio2017}) and in starburst galaxies is usually in excess of what the true `cold' gas component predicts~\citep{Kamenetzky2014}. But although there is probably some overlap between the J = $6-5$ and J = $3-2$ emitting gas, in bulk they preferentially probe two different temperature regimes.

We therefore generate separate models for a `warm' component matching the J = $6-5$ profile and a `cool' component matching the J = $3-2$ line. Models of the two temperature components have the same velocity structure, gas inclination, dust temperature and position angle. The gas mass, density and temperature distribution are different for the two models. In effect we require the cool and warm gas to have the same morphology, but permit the gas' extent and density to vary within that morphology.

Due to the simpler nature of the CO J = $6-5$ line profiles compared to those of CO J = $3-2$, we first use CO J = $6-5$ observations to constrain the velocity structure, inclination and position angle, and dust temperature. We then applied the fiducial fit parameters from the CO J = $6-5$ model to find the gas density distribution and gas temperature of CO J = $3-2$.

The model parameters are found by eye due to the complexity of the nuclear gas structure and dynamics, the large parameter space, and the number of fit parameters available to the model. The parameters are chosen to minimize the apparent differences between the modeled line profiles and the observed line profiles extracted at points separated by a beam FWHM along the major and minor axes of the modeled gas. Once a fiducial fit is found, ranges on acceptable fit parameters are found by varying one parameter at a time until the model's line profiles are no longer acceptably close to the observed line profiles. ``Acceptable'' fits are determined by a combination of factors: first, the modeled line profiles on average cannot appear to differ from the observed line profiles by more than $\sim$ 25$\%$ in height, width, or central velocity. 
The fits focus on the brightest region along the gas' major axis.
Second, the shape of the modeled line profiles should approximately match that of the observed line profiles.

Many of the parameters can be well constrained by the observed line profiles, for example $r_{sh}$ by the rate of decay of the line profile heights with radius and $v_{turb}$ by their widths. The singly peaked profiles and lack of gradient in their central velocities constrains the inclination to nearly face-on models. Inclinations between 45$^{\circ}$ and 135$^{\circ}$ are strongly disfavored because they showed doubly peaked profiles. Other parameters are more difficult to constrain. $v_{circ}$ and $r_{circ}$ are the least constrained because they are anti-correlated (see Equation~\ref{eq:v}) leading to large fit ranges as they trade off against each other. However, $v_{circ}$ is not wholly unconstrained. It cannot be so high as to shift the line centers, but must be high enough to create the singly peaked profiles. The temperature $T$ is constrained by the brightness of the lines, with higher temperatures (and therefore higher CO rotational excitation) corresponding to brighter lines.

\begin{figure}[ht!]
	\begin{centering}
		\includegraphics[width=0.9\linewidth]{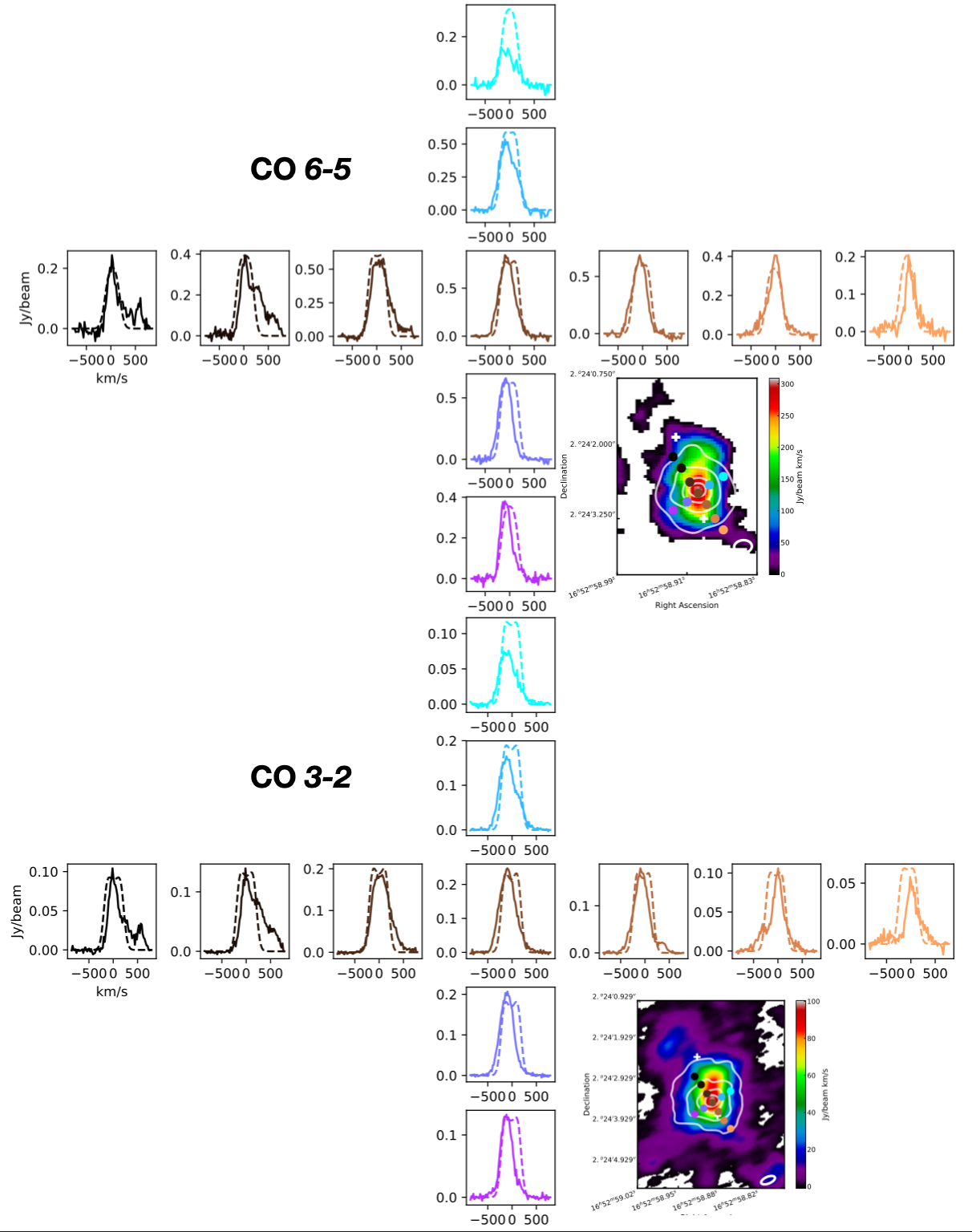}
	\caption{The line profiles for the fiducial fit CO J = $6-5$ (\textbf{top}) and J = $3-2$ (\textbf{bottom}) LIME models compared to the observed CO J = $6-5$ and J = $3-2$ line profiles extracted along 34$^{\circ}$, the angle between the maximum of the highly blueshifted and redshifted gas. \textit{Dashed Lines:} model line profiles at the location color-coded in the inset image. \textit{Solid Lines:} Observed line profiles at the location color-coded in the inset image. \textbf{Inset Image:} The color map is the moment 0 map for the corresponding observation, and contours show the model's integrated line emission.} \label{fig:best fit lines} 
	\end{centering}
\end{figure}

All available parameters, fit values, and acceptable ranges of model parameters are reported in Table~\ref{table:disk values}. The fiducial fit model whose emission profiles most closely match the observed profiles is a nearly face-on pancake-like distribution of gas, with a scale height $z_{sh}$ of 20-60 pc and a radial scale length of 390 pc at an inclination of 22.5$^{\circ}$ from face-on. The line profiles extracted from these models along the semi-major axis of the gas concentration for CO J = $6-5$ and J = $3-2$ are compared to those from the data in Figure~\ref{fig:best fit lines}.

\begin{deluxetable}{ c|ccc  }
	\label{table:disk values} 
		\tablecaption{Fiducial LIME model values for CO J = $3-2$ and J = $6-5$ (\textit{column 2}), as well as the ranges in parameter space that resulted in acceptable models (\textit{column 3}). Velocity structure, gas inclination, and dust temperature is shared between the two models while gas mass, density distribution, and gas temperature vary between the two. Gas masses are calculated by integrating the modeled density over the observed extent of the gas (the high signal-to-noise ratio areas in Figures~\ref{fig:CO32 moments} and~\ref{fig:CO65 moments}).  }	
\tablehead{ \colhead{Variable Name} & \colhead{Fit Value} & \colhead{Range} }  
	
\startdata
			\multicolumn{3}{c}{Shared Fiducial Model Values} \\
			Inclination & 22.5$^{\circ}$ & $0^{\circ} - 45^{\circ}$, $135^{\circ} - 180^{\circ}$\\
			Position Angle & 33.7$^{\circ}$ & $30^{\circ} -  40^{\circ}$\\
			T$_{dust}$ & 30 K & $20$~K $-~100$~K \\ 
			v$_{turb}$ & 140 km/s & $100$~K~$-$~$160$~km/s \\ 
			v$_{circ}$ & 100 km/s & 50 km/s $-$ 300 km/s \\ 
			r$_{circ}$ & 200 pc & 0 pc $-$ 500 pc\\ 
			\multicolumn{3}{c}{CO J = $6-5$ Fiducial Model Values} \\
			Gas Mass & 6.0$\times$10$^{8}$  M$_{\odot}$ & $4.0\times10^{8} - 7.8\times10^{8}$ M$_{\odot}$\\
			r$_{sh}$ & 390 pc & 240 pc $-$ 500 pc \\ 
			z$_{sh}$ & 20 pc & 18 pc $-$ 60 pc\\
			$a_{n}$ & 1.0 & 0.6 $-$ 1.0 \\ 
			T & 2,000 K & 1,500 K $-$ 3,000 K \\ 
			\multicolumn{3}{c}{CO J = $3-2$ Fiducial Model Values} \\
			Gas Mass & 7.6$\times$10$^{8}$  M$_{\odot}$ & $2.9\times10^{8}$ M$_{\odot} - 9.2\times10^{8}$ M$_{\odot}$\\
			r$_{sh}$ & 350 pc & 250 pc $-$ 400 pc \\ 
			r$_{sh, x \textless 0}$ & 250 pc & 200 $-$ 400 pc \\ 
			z$_{sh}$ & 13 pc & 10 pc $-$ 16 pc\\ 
			$a_{n}$ & 1.0 & $0.8 - 1.8$ \\ 
			T & 600 K & 400 K $-$ 1500 K \\ 
\enddata
\end{deluxetable}

\begin{figure}[ht!]
	\begin{centering}
		\includegraphics[width=0.75\linewidth]{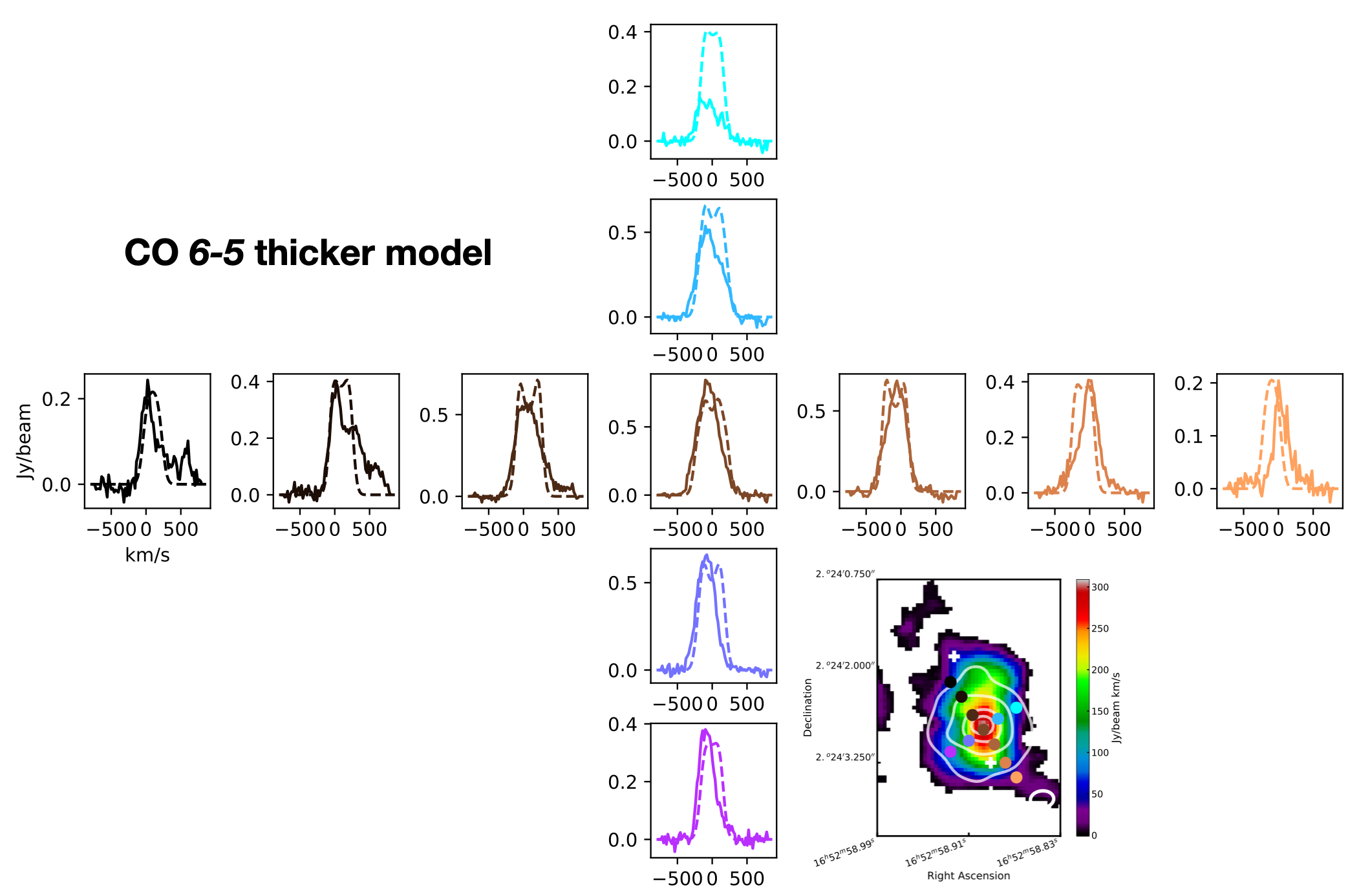}
		\caption{The line profiles for the thicker fiducial fit CO J = $6-5$ LIME model compared to the observed CO J = $6-5$ line profiles extracted along 34$^{\circ}$, the angle between the maximum of the highly blueshifted and redshifted gas. The model parameters of the gas concentration are reported in Table~\ref{table:thick disk values}. \textit{Dashed Lines:} model line profiles at the location color-coded in the inset image. \textit{Solid Lines:} Observed line profiles at the location color-coded in the inset image. \textbf{Inset Image:} The color map is the moment 0 map for the corresponding observation, and contours show the model's integrated line emission.  }
		\label{fig:best fit thick model lines}
	\end{centering}
\end{figure}

A wide range of parameter space resulted in fits of similar quality to our fiducial model.  As an example, we include a thicker-disk model with the parameters in Table~\ref{table:thick disk values} with the corresponding modeled line profiles in Figure~\ref{fig:best fit thick model lines}. This model is thicker than the models presented in Table~\ref{table:disk values}, with a vertical scale height $z_{sh}$ of 60 pc instead of 20 pc, a gas concentration that is more horizontally concentrated ($r_{sh}$ = 270 pc instead of 390 pc), and with a lower ratio of turbulent to circular velocity ($v_{turb} / v_{circ}$ = 0.44 instead of 1.4). While this thicker model's parameters are all within the thin disk's acceptable ranges and still capture the majority of the observed emission (line profile width, intensity, and in some cases line profile shape), the line profiles are more doubly peaked and less representative of the observed line profiles.

\begin{deluxetable}{ c|cc}
	\label{table:thick disk values}
			\tablehead{ \colhead{Variable Name} & \colhead{Thick Disk Fit Value} & \colhead{Thin Disk Fiducial Fit Value} }
			\tablecaption{Fiducial LIME model values for the thicker modeled CO J = $6-5$ gas concentration compared to the thinner model. Parameters not listed are the same as in Table \ref{table:disk values}. The thick model's line profiles are presented in Figure~\ref{fig:best fit thick model lines}. The thin model's values are taken from Table~\ref{table:disk values} and the line profiles are presented in Figure~\ref{fig:best fit lines}. }
\startdata
			z$_{sh}$ & 60 pc & 20 pc\\
			r$_{sh}$ & 270 pc & 390 pc\\ 
			v$_{turb}$ & 110 km/s &  140 km/s \\ 
			v$_{circ}$ & 250 km/s & 100 km/s \\ 
			r$_{circ}$ & 10 pc & 200 pc \\ 
			Gas Mass & 7.0$\times$10$^{8}$  M$_{\odot}$ & 6.0$\times$10$^{8}$  M$_{\odot}$ \\
			T & 2,500 K & 2,000 K \\ 
\enddata
\end{deluxetable}

\subsection{Fiducial Model Parameters in Context}\label{sec:model physicality}

We place the model parameters in a physical context by comparing them to the findings of~\cite{Kamenetzky2014} who use \textit{Herschel Space Observatory} and other supplementary observations (see their Table 16) of CO J = $1-0$ to $13-12$ to model gas temperatures, H$_2$ column densities, gas masses, and H$_2$ number densities for NGC 6240. We find our CO J = $3-2$ model has a temperature and mass consistent with their calculations. The CO J = $6-5$ model is on the high end of both temperature and mass, but the total modeled mass of the gas is consistent with other findings for NGC 6240.

Our models indicate the CO J = $6-5$ emission is dominated by  relatively hot (T = 2,000 K) gas, and the CO J = $3-2$ emission by warm (T = 600 K) gas.
\cite{Kamenetzky2014} modeled the CO emission as a combination of cold (16 K with 1$\sigma$ range [5, 50] K) and hot (1,260 K with 1$\sigma$ range [790, 2,000] K) gas. They found the CO J = $3-2$ emission is dominated by a mixture of both cold and hot  gas, while the CO J = $6-5$ emission is produced primarily by the hot gas component. Our CO J = $3-2$ fiducial model temperature of 600 K lies between the temperatures of~\cite{Kamenetzky2014}'s cool and hot components, supporting the argument that the CO J = $3-2$ emission is created by a combination of gas temperatures. Also in line with their model, our model finds the CO J = $6-5$ emission to be dominated by hot (2,000 K) gas, at the upper bound of their modeled 1$\sigma$ range on gas temperature. 
Our model temperatures may be on the higher end due to the much smaller ALMA beam sizes that probe the nuclear region while~\cite{Kamenetzky2014}'s models incorporate \textit{Herschel} data that combines emission from both nuclear and extended regions. The nuclear region is likely to be much warmer than the extended gas due to star formation and AGN activity concentrated in the central kpcs.
The high gas temperatures are supported by the presence of prominent H$_2 1-0 S(1)$ emission that requires temperatures $\textgreater$ 1,000 K to be excited (\cite{Max2005};~\cite{Meijerink2013}). 

The model fit dust temperature range of [20, 100] contains~\cite{Kamenetzky2014}'s best fit dust temperature of 56 K that we used in our continuum mass calculations (Section~\ref{sec:mass}).

~\cite{Kamenetzky2014} also models the masses of these cool and warm components, finding a cool gas mass of 2$\times$10$^{9} \pm$ 5$\times$10$^{8}$ M$_{\odot}$ and a warm gas mass of 4$\times$10$^{8} \pm$ 10$^{8}$ M$_{\odot}$. Our total modeled mass for CO J = $3-2$ is 7.6$\times$10$^{8}$ M$_{\odot}$, higher than~\cite{Kamenetzky2014}'s modeled warm gas mass but lower than their cool gas mass. This is again consistent with the theory that CO J = $3-2$ is a mixture of cool and warm gas components. The ranges of CO J=$6-5$ gas mass modeled by LIME  spans a factor of two (Table~\ref{table:disk values}) that contain~\cite{Kamenetzky2014}'s masses. However, our modeled hot gas mass (captured by simulations of CO J = $6-5$) is 6$-$7$\times$10$^8$ M$_{\odot}$, higher than their best-fit warm gas mass by around 50$\%$. 
Despite the high mass of the hot gas in the model, the total modeled mass of $\sim$1.4$\times$10$^{9}$ M$_{\odot}$ (CO J = $3-2$ plus CO J = $6-5$) is close to the nuclear region's mass calculated from the continuum in Section~\ref{sec:mass} (1.2$\times$10$^{9}$ M$_{\odot}$, region 1, Figure~\ref{fig:mass regions}). 
This presents a possible conundrum: the dust continuum observations in Section~\ref{sec:mass} suggest that much of the dust emission is not captured by these observations, while the modeled gas masses do appear to capture the majority of the gas mass. It is possible that this discrepancy comes down to the high $L_{CO}$/$L_{FIR}$ ratio in NGC 6240, around 10 times higher than other nearby galaxies~\citep{Kamenetzky2014}. The CO emission could be dominated by the shocked gas observed with ALMA, while there is more dust and cold (less luminous) gas outside of the ALMA observations that would included in \textit{Herschel} observations.  In other words, due to the concentrated nuclear excitation of CO in NGC 6240, the ALMA observations efficiently captured CO emission, but not the extended dust emission.

\subsection{Transience and Stability of Fiducial Models}\label{sec:velocities and transience}
In this section we examine the modeled velocity dispersion and circular velocities in the context of the model geometry, finding that the model suggests a transient structure that is stable against collapse.

The ratio of our average rotational velocity to our modeled velocity dispersion is $\textless v \textgreater$/ $\sigma \sim$ 0.7 for the thin fiducial models presented in Figure~\ref{fig:best fit lines}.  According to~\cite{Tacconi1999} this is in the range that indicates a disk that must be geometrically thick. However, the fiducial models are quite thin, with $z_{sh}$ of only 20 pc and $r_{sh}$ of 250$-$390 pc. This aspect ratio is unlikely for a self-gravitating structure with such high velocity dispersion since the dispersion would have the effect of ``puffing up'' the disk.  It is also possible that with such a thin concentration of gas, the modeled high velocity dispersion could partially be accounted for by shear. That is, the gas concentration could be undergoing tidal disruption in the plane of the sky.

The thicker fiducial model presented in Table~\ref{table:thick disk values} and Figure~\ref{fig:best fit thick model lines} has a higher ratio of $\textless v \textgreater$/ $\sigma \sim$ 2.3 and a thicker geometry ($z_{sh}$ = 60 pc, $r_{sh}$ = 270 pc). In both senses this model is less extreme than the thinner model, though it visibly does not fit the observed line profiles as well. That the thicker model would correspond to a lower velocity dispersion may seem counterintuitive, expecting that higher velocity dispersions would ``puff up" the gas. However, if one invokes shear in the plane of the sky this would artificially inflate the velocity dispersion for thinner and more optically thin gas concentrations.

We can also conduct a back-of-the-envelope calculation to determine the lifetime of the gas concentration given its modeled velocity dispersion and vertical extent. For the thin model, the velocity dispersion is 140 km/s (assuming no contribution from shear) and the vertical extent of the concentration is $\sim$ 40 pc (two vertical scale heights). Gas moving at 140 km/s would travel this distance after 300,000 years. This is an extremely short timescale in the context of galaxy mergers that occur over timescales of Gyr, suggesting the model represents a highly transient structure. For the thicker model, gas traveling at a speed equal to the velocity dispersion of 110 km/s would travel two scale heights' distance (120 pc) in $\sim$ 1 Myr, still a short timescale in the context of galaxy mergers. Such short lifetimes are unlikely for a self-gravitating structure.

As a further test of the models in the context of a self-gravitating structure, we can compare the velocity dispersion and circular velocities to the escape velocity from the modeled gas concentration. For this calculation, we use the escape velocity of $\sqrt{2 G M / r}$ for the entire gas concentration with $M = $1.4$\times$10$^{9}$ M$_{\odot}$. At the radial scale height of the thinner model $r_{sh}$ = 390 pc, $v_{esc}$ is 170 km/s. This value is similar to the thin models' velocity dispersion of 140 km/s, and is less than the circular velocity (100 km/s) added to the modeled dispersion. At the radial scale height of the thicker model $r_{sh}$ = 270 pc, $v_{esc}$ is 210 km/s. This exceeds the thicker model's turbulent velocity of 110 km/s but is less than its circular velocity alone ($v_{circ}$ = 250 km/s). This means both models' gas velocity can exceed the escape velocity of the gas concentration, indicating it is extremely unlikely this gas is a self-gravitating disk.

Given the fiducial LIME model, we can also check the stability of the gas concentration by calculating the Toomre parameter $Q$ = $\sigma_r \kappa $ / $\pi$ G $\Sigma_{gas}$. Here $\kappa$ = $\sqrt{3}$ $v_{max}$ / $R$ is the epicyclic frequency, $\sigma_r$ is the line-of-sight velocity
dispersion, and $\Sigma_{gas}$ is the mass surface density of the gas~\citep{Toomre1964}. $Q$ for the thinner LIME model is 2.3 for CO J = $6-5$ and 2.5 for the CO J = $3-2$ model. A $Q$ $\textgreater$ 1 means the gas is stable against collapse at this time, with the majority of models within the acceptable parameter ranges fitting into this category. These high values of $Q$ are consistent with this highly turbulent system. The majority of the luminosity being generated in this central mass concentration is then unlikely to be due to star formation, consistent with other papers that argue for heating from shocks and superwinds originating in starbursts around the nuclei, not in the central region between the two nuclei (e.g.~\cite{Tecza2000};~\cite{Max2005};~\cite{Engel2010}).

\subsection{Model Residuals: What is Captured by the Model?}~\label{sec:resid}

\begin{figure}[ht!]
	\plotone{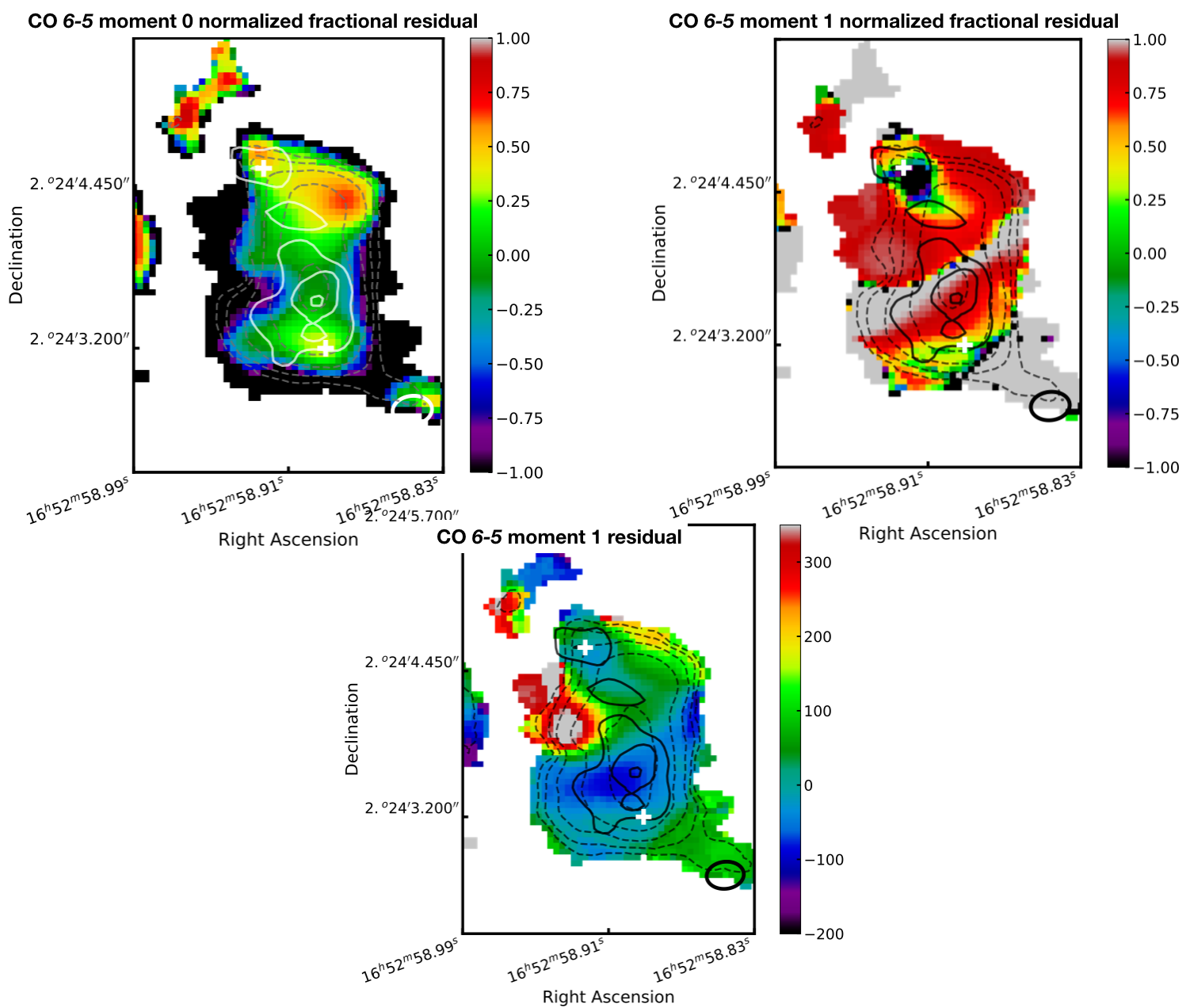}
	\caption{Normalized fractional residuals (\textit{color}) between CO J = $6-5$ data and simulation for moment 0 (\textit{upper left}) and moment 1 (\textit{upper right}). The \textit{lower panel} shows the non-normalized residual in moment 1. \textit{Dashed contours} correspond to the moment 0 data, \textit{solid contours} correspond to the 678 GHz continuum emission, and \textit{crosses} show the locations of the two AGN. Beam FWHM contours are shown in the bottom right of each panel.  \label{fig:CO65_residuals} }
\end{figure}

\begin{figure}[ht!]
	\plotone{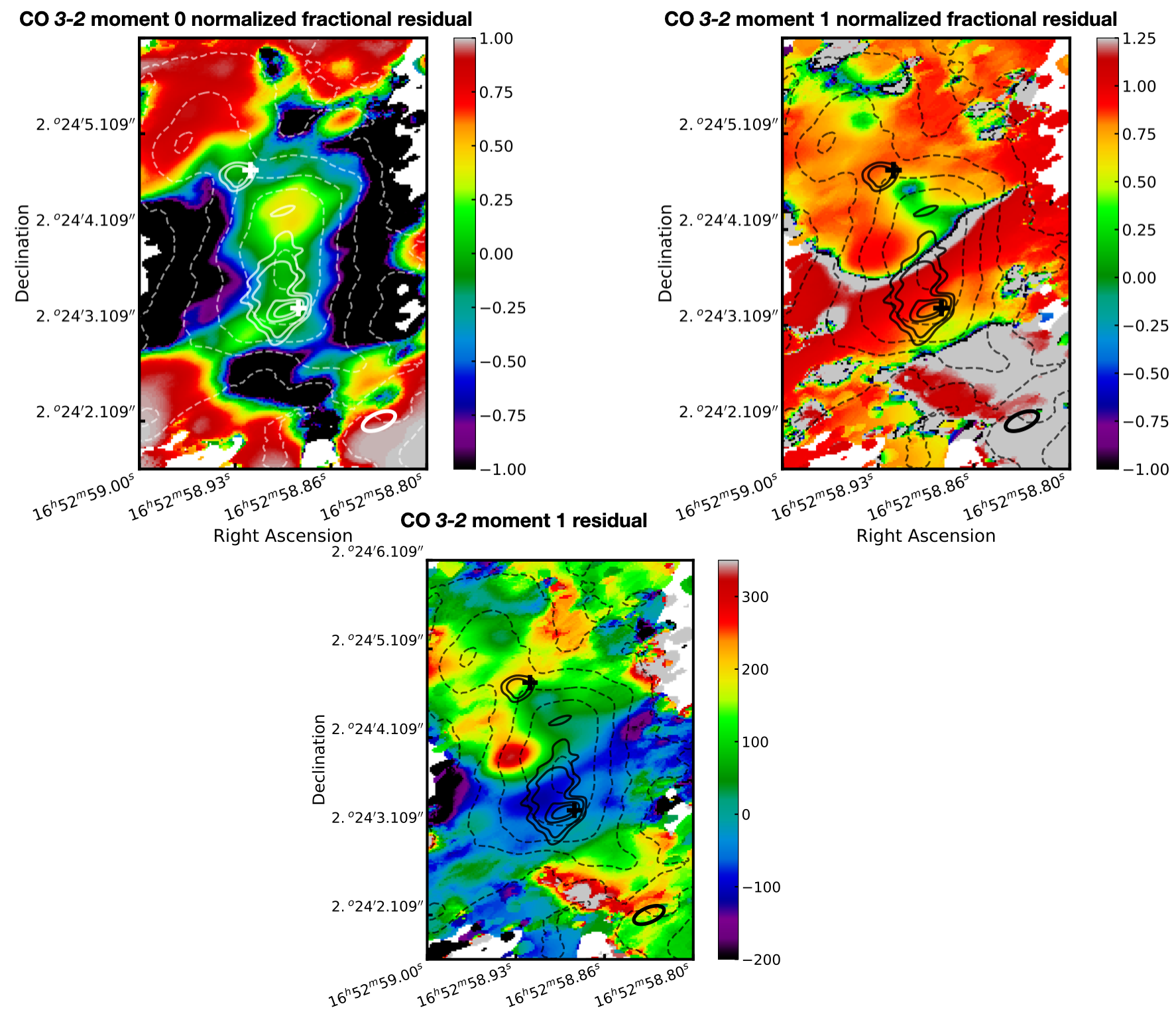}
	\caption{Normalized fractional residuals (\textit{color}) between CO J = $3-2$ data and simulation for moment 0 (\textit{upper left}) and moment 1 (\textit{upper right}). The \textit{lower panel} shows the non-normalized residual in moment 1. \textit{Dashed contours} correspond to the moment 0 data, \textit{solid contours} correspond to the 345 GHz continuum emission,  and \textit{crosses} show the locations of the two AGN. Beam FWHM contours are shown in the bottom right of each panel.  \label{fig:CO32_residuals} }
\end{figure}

Figures~\ref{fig:CO65_residuals} and~\ref{fig:CO32_residuals} show the moment 0 and 1 normalized residuals (upper panels) and moment 1 absolute residual (lower panel) between ALMA data and the thinner fiducial LIME models for CO J = $6-5$ and J = $3-2$ (presented in Table~\ref{table:disk values} and Figure~\ref{fig:best fit lines}). The normalized fractional residuals are calculated by subtracting the model from the observation and dividing by the observation for each pixel. 
 We present the residuals for the thinner and not the thicker models because they most closely represent the observed line profiles.

The residuals for each moment take a similar shape for both emission lines.  The low values of the moment 0 fractional residual in the brightest central region indicate that the test model captures the majority of the line emission in this central region. However, the  model misses the northernmost portion of the observed gas concentration, especially apparent in the CO J = $6-5$ residual. The test model does not accurately describe the $\textless$v$\textgreater$ data, as evidenced by the large values and amount of structure that remains in the fractional residuals of moment 1 for both CO J = $6-5$ and J = $3-2$. The non-normalized residual of $\textless$v$\textgreater$ (observed - model), shown in the bottom panel of each figure, remains largely unchanged from the observed values. This indicates that the model does not reproduce the velocity structure well at all, and explains why the normalized residuals are all close to one. 

The residuals shed light on what components of NGC 6240's molecular gas are captured by the test model, if any. There are aspects of the gas that are clearly not captured, for example the modeled line profiles in Figure~\ref{fig:best fit lines} miss a shoulder of highly redshifted emission ($\textgreater$ 300 km/s) in the northeast plotted therein as dark brown and black line profiles. This emission corresponds to the highly redshifted gas visible in the average velocity maps ($\sim$ 300 km/s in Figure~\ref{fig:CO32 moments} and $\sim$ 400 km/s in Figure~\ref{fig:CO65 moments}) and in the channel maps (up to 665 km/s in Figure~\ref{fig: CO32_channelmap} and up to 724 km/s in Figure~\ref{fig: CO65_channelmap}). It also corresponds to the highly redshifted gas in Figure~\ref{fig:redblueshifted} with an average velocity of $\sim$400 km/s when isolated from the quiescent gas, or $\sim$250 km/s when including the quiescent gas. 
This gas is dim compared to the majority of the line emission, and as such does not augment the normalized fractional residual moment 0 maps above $\sim$0 despite not being described by the models.

Despite missing this high velocity gas, the low values for the integrated line emission residuals in the central/southern portion of the central gas concentration indicate that the models capture the majority of the CO emission in this, the brightest and most massive part of the ALMA observations. We calculate that the model captures 96$\%$ of the observed emission, a value found by dividing the sum of the modeled CO J = $6 - 5$ emission divided by the sum of the observed emission within the 5$\sigma$ contour area.  Capturing the majority of the emission indicates we accurately model the gas mass and gas temperature, which are degenerate parameters that each increase total emission when increased. Constraints on gas mass and temperature are provided by the modeled line profile shapes with exceedingly high masses or low temperatures resulting in doubly-peaked profiles. Accurate models of mass and temperature do not directly imply accurate models of gas distribution. The other parameters important for this are $r_{sh}$, well constrained by the visible extent of the gas, and $z_{sh}$. $z_{sh}$ is somewhat difficult to constrain because the gas is nearly face-on, but thick gas results in doubly peaked profiles constraining $z_{sh}$ to small values (tens of parsecs). Density is directly calculated from the gas mass, $r_{sh}$, and $z_{sh}$. Our constraints on these values and the gas temperature indicate we have successfully modeled the \textit{physical distribution} and density of the gas. Our models found that this central gas region is pancake-like: quite thin (tens of pc) with respect to its extent (hundreds of pc).

While the fractional residuals of the line emission appear to accurately capture the \textit{distribution} of the gas, the velocity residuals tell a different story. The normalized fractional velocity residuals (upper right panel of each figure) are comprised almost entirely of values close to one, with absolute residuals (lower panel of each figure) close to the average velocities of the observed gas. These large residuals suggest the simple exponentially decaying velocity profile (like that of a disk) does not accurately capture the \textit{kinematics} of the gas. When taken with the arguments in Section~\ref{sec:model physicality} that the modeled velocity dispersion indicates a transient structure whose velocities can exceed the escape velocity of the gas concentration, we suggest that the nuclear molecular gas emission is dominated by a thin pancake-like gas concentration that is not rotating like a disk. This conclusion is also consistent with the findings of Section~\ref{sec:model physicality}, finding our masses and temperatures fit within current observations but the high velocity dispersion relative to the circular velocity is unlikely for such a thin geometry if the gas were to be a self-gravitating disk.

\section{Is the Central Molecular Gas a Tidal Bridge?}\label{sec:tidal bridge}

\begin{figure}[h!]
	\begin{centering}
		\includegraphics[width=0.9\linewidth]{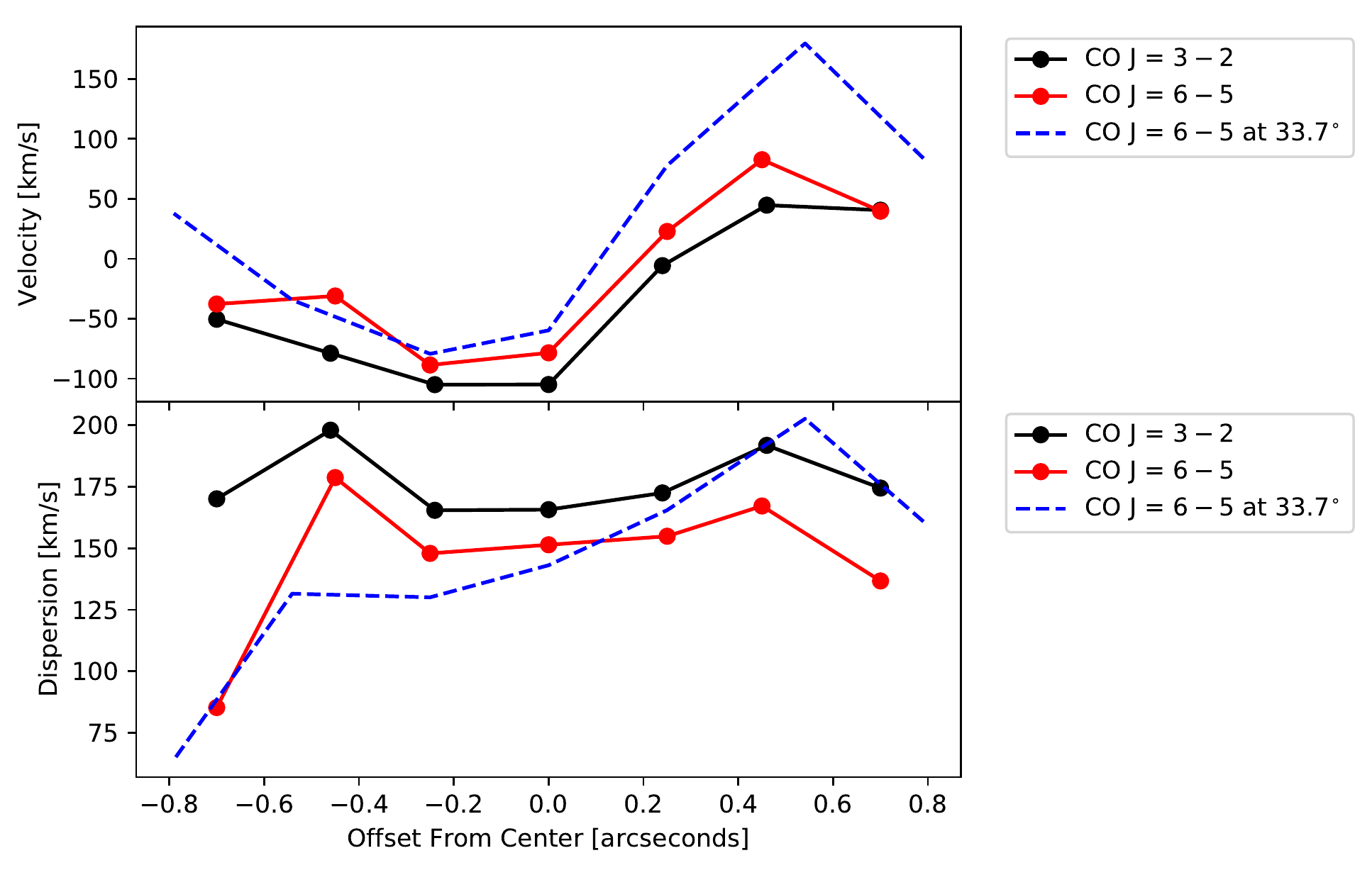}
		\caption{The average velocity and dispersion of CO $6-5$ (red) and CO 3-2 (black) extracted along the semimajor axis of the central gas region (0$^{\circ}$ \textit{solid}) and along the axis of the modeled gas (33.7$^{\circ}$ \textit{blue dashed}).  }
		\label{fig:CO65_mom1_mom2_extraction}
	\end{centering}
\end{figure}

The models indicate that the gas distribution has a vertical thickness of tens of parsecs with a horizontal extent of hundreds of parsecs. At the location of the brightest molecular gas emission the fractional moment 0 residuals between the observations and the models are close to zero. Capturing the majority of the emission indicates we accurately model the gas mass and gas temperature, which are degenerate parameters that each increase total emission when increased. However, as argued in Section~\ref{sec:resid}, the molecular gas kinematics are not well captured by the disk model as demonstrated by the high fractional and absolute residual values for $\textless v \textgreater$ (moment 1). Additionally, our calculations and discussion in Section~\ref{sec:velocities and transience} indicate the gas is unlikely to be a self-gravitating disk, consistent with arguments from~\cite{Treister2020},~\cite{Engel2010} and~\cite{Cicone2018} mentioned in the introduction. The contra-indication of a gravitationally stable disk is true for both the thin and thicker fiducial models. Therefore, we must consider other possibilities to explain the molecular gas kinematics. The extended distribution of the model could suggest a tidal bridge between the two nuclei, stretched by tidal forces along the semi-major axis of the observed CO emission at PA = 0$^{\circ}$. Other scenarios are possible, such as gas infall and outflow and multiple disks.  However, it would not be surprising for there to be tidally disruption of dissipative gas in the center of two merging galaxies that have recently undergone a first pass.  Furthermore, NGC 6240 does not exhibit the clear evidence for multiple disks as in other double-nucleus galaxies, such as Arp 220~\citep{Wheeler2020}.  Given that there is already evidence for a tidal bridge between the nuclei, we explore this interpretation.

If the gas is indeed a tidal bridge, its low density means it is unlikely to be in virial equilibrium and could result in sub-thermal excitation.  The X-factor (the relationship between $N(H_2)$ and the CO luminosity) only applies for virialized, denser clouds, and can be radically modified for non-virialized, thin clouds~\citep{Tacconi2008}. Physically this can manifest as higher than expected CO line luminosity compared to the observed H$_2$ and dust luminosity as well as offsets between the peak of CO line emission and the peak of H$_2$ and dust emission \citep{Engel2010, Tacconi2008}. Our observations show an offset between CO and dust emission peaks, and comparison to H$_2 (1-0) ~S(1)$~\citep{Max2005} in Figure~\ref{fig:Max2005 comparison} also shows an offset between H$_2$ and CO emission. This sub-thermal and therefore optically thinner emission could also help to explain the anomalously high $L_{CO}/L_{FIR}$ observed in NGC 6240, around 10 times higher than other nearby galaxies~\citep{Kamenetzky2014}. To check for the possibility of sub-thermal emission we compare the observed density to the critical densities $n_{crit}$ for the observed lines. The modeled average $H_2$ density is $\sim$ 1$\times$10$^{3}$ cm$^{-3}$ for both the thicker and thinner fiducial models. The critical density $n_{crit}$ for CO J = $3-2$ emission of 3.6$\times$10$^{4}$ cm$^{-3}$ and  2.9$\times$10$^{5}$ cm$^{-3}$ for CO J = $6-5$. $n_{crit}$ for both transitions is 1-2 orders of magnitude higher than the modeled density, indicating that sub-thermal emission is likely in this central region.  

The presence of a tidal bridge or ribbon between the two nuclei is also supported by observations of H$_2$ $1-0$ S(1) and S(5) presented in~\cite{Max2005} and recreated in our Figure~\ref{fig:Max2005 comparison}. They observe a ribbon of H$_2$ with a reverse S shape (their Figure 10) extending between the northern and southern nuclei and postulate it could be material flowing along a bridge connecting the two progenitor galaxy nuclei. They compare this geometry to the tidal bridges predicted by computer simulations (e.g.~\cite{Barnes1991}; see also Figs. 9 and 10 in~\cite{Barnes1996}; Figs. 1 and 4 in~\cite{Barnes2002}). One important note is that these simulations show tidal bridges on scale of $\sim$ 5-40 kpc, while the projected separation between the nuclei in NGC 6240 is only $\sim$ 1 kpc. 
Nonetheless, smaller scale simulations (e.g.~\cite{Hopkins2013}) also show significant mass flowing between galaxy nuclei on scales $\lesssim$ 5 kpc. 
While the irregular H$_2$ morphology does not follow the morphology of CO, this could be explained by an altered CO-to-H$_2$ conversion factor in this nuclear region~\citep{Tacconi2008} causing CO to be brightened with respect to H$_2$ and dust along the N/S axis as described above. The differences in the two observations are unlikely to be caused by differences in sensitivities or beam sizes, as the H$_2$ observations show offsets in peak brightness located at the southern nucleus, farther than 1.5 beam FWHM away from the peak brightness observed in the CO observations.

Tidal bridges and filaments have been shown to have relatively small line-widths ($\sim$ 50 -- 100 km/s) in tidal dwarf galaxies~\citep{Braine2001}. In larger galaxies the linewidths and velocities are larger, for example the CO FWHM  is $\sim$ 200 km/s with $\textless v \textgreater \sim$ 130 km/s in the bridge between the Taffy galaxies~\citep{Braine2003}. In Arp 194 (total dynamical mass $\textgreater$ 10$^{11}$ M$_{\odot}$), the dispersion is $\sim$ 25-125 km/s with $| \textless v \textgreater | \textless$ 150 km/s in the bridge connecting the two galaxies (\cite{Zasov2016}, their Figure 3). In Figure~\ref{fig:CO65_mom1_mom2_extraction} we plot the velocity and dispersion along the major axis of the observed gas (PA = 0$^{\circ}$) and the major axis of the modeled gas (PA = 33.7$^{\circ}$) to compare to these observed line widths and velocities. For both extractions, the velocities all lie below approximately $|$100$|$ km/s, with a structure that moves smoothly from redshifted to blueshifted, as expected for a tidal bridge. The total velocity dispersion shows values of 150-180 km/s, slightly higher than the bridge in Arp 194 but within range of the Taffy galaxy bridge. Within the context of these observations a tidal bridge remains a plausible explanation of the internuclear molecular gas.

\section{Possible Fates of the Molecular Gas}\label{sec:merger sims}

The geometry of the merger that formed NGC 6240 is discussed in detail in~\cite{Engel2010}. They propose that the merger has a geometry that tends towards coplanar/prograde because of the extended tidal tails observed in NGC 6240. 
~\cite{Lotz2008} presents simulations of equal-mass gas-rich mergers with a variety of geometries, including coplanar prograde encounters like NGC 6240. All coplanar encounters of gas-rich Sbc-type galaxies simulated in~\cite{Lotz2008} have two peaks of star formation, one around 1 Gyr and another stronger starburst around 2 Gyr after the merger begins. Consistently, ~\cite{Tecza2000} argued that we are observing NGC 6240 shortly after the first encounter triggered an initial starburst, as did~\cite{Engel2010} who argued NGC 6240 ``has currently elevated levels of star formation compared to a quiescent galaxy; and will experience another, likely stronger, peak in star formation rate in the near future when the galaxies coalesce". 
~\cite{Engel2010} also make an important note that NGC 6240 is currently barely below the ULIRG classification of L$_{IR} \gtrsim$ 10$^{12}$ L$_{\odot}$, and will likely breach that threshold once the second starburst is triggered, supported by the stronger second starburst predicted by the simulated merger models in~\cite{Lotz2008}. 

It is possible that the tidal bridge will fall onto the nuclei prior to the second pass and final coalescence, possibly triggering the second starburst or feeding further AGN activity. We calculate the free-fall time of the gas of mass $m$ onto a nucleus of mass $M$ using 
\begin{equation}\label{eq:tff}
t_{ff} = \frac{\pi}{2}\frac{R^{3/2}}{\sqrt{2G(M+m)}} ,
\end{equation}
where $R$ is the distance between the gas and the nucleus. As this is an order-of-magnitude calculation, we split the molecular gas mass in half and assume half will fall to the northern nucleus and half to the southern. We use the nuclei masses of 1.3$\times$10$^{10}$ M$_{\odot}$ for the southern nucleus and 2.5$\times$10$^{9}$ M$_{\odot}$ for the northern~\citep{Engel2010}. The projected radius from the maximum of the integrated line emission to the northern nucleus is 590 pc and 145 pc to the southern. This means $t_{ff, northern}$ = 4.3$\times$10$^{6}$ years and $t_{ff, southern}$ = 2.5$\times$10$^6$ years. That is, the free-fall time of the gas onto the nuclei is a few Myr while the time between the first pass and the  galaxies' maximum separation prior to the second pass is likely to be $\sim$ 400 Myr~\citep{Lotz2008}. Therefore, it is likely that this gas will fall onto the nuclei prior to the next pass, possibly adding to the current nuclear starburst. One issue with this interpretation is the high modeled velocity dispersion and small vertical extent that indicate a transient structure that will dissipate after $\sim$0.3 Myr (see argument at the end of Section~\ref{sec:model physicality}). Therefore, it is possible that the nuclear concentration of gas will dissipate prior to streaming onto the nuclei.

\section{Summary}\label{sec:summary}

NGC 6240 presents an interesting test case of a galaxy merger between first pass and final coalescence, an intermediate and turbulent stage of galaxy evolution. It provides a detailed example to study what comes just before galaxies evolve into ULIRGs, as it is just below the classification threshold of a ULIRG but is likely to evolve into a ULIRG when the second, stronger, merger-induced starburst is triggered.	We presented high-resolution ALMA observations of CO J = $6-5$ and J = $3-2$, the first observations of the nuclear region of NGC 6240 in CO J = $6-5$ and the highest resolution observations to date in CO J = $3-2$. We observe similar morphology to previous CO observations, notably, a concentration of gas between the two nuclei that is distinct from the continuum that is itself centered around the nuclei \cite[etc.]{Tacconi1999, Scoville2014}. 
We model the molecular gas density and velocity distributions using LIME and find a thin, pancake-like distribution of gas whose velocities and velocity dispersions indicate a transient concentration that is unlikely to be a self-gravitating disk. The model captures the majority of the gas' emission but fails to capture the gas' kinematics, as demonstrated by the residual maps.
We instead argue that the nuclear region observation is consistent with superposed emission from a tidal bridge and highly redshifted gas. This work demonstrates the importance of high-resolution multi-line observations when trying to disentangle the effects of energetic gas acceleration mechanisms, star formation, and tidal forces in the central regions of major mergers. 

We argue that the majority of the central molecular gas concentration is a tidal bridge connecting the two nuclei of the progenitor galaxies. 
Our fiducial models show that this central gas region is likely quite thin (vertical scale height of 20 to 60 pc) with respect to its extent (horizontal scale height of 240 to 500 pc). That this central gas is a bridge connecting the nuclei is supported by~\cite{Engel2010} who argue for an altered CO-to-H$_2$ conversion factor in the central region that would be exacerbated by a drawn-out, thin, and diffuse gas bridge. The H$_2$ observations presented in~\cite{Max2005} also support the idea of a tidal bridge, which are in turn motivated by simulations from~\cite{Barnes1991} and~\cite{Barnes1996} that show mergers can result in material flowing along bridges connecting progenitor galaxies. Kinematic arguments from~\cite{Treister2020} suggest the gas bridge connects the two nuclei with clear streaming kinematics.
Gas that is subject to gravitational torques, such as that in a tidal bridge in the nuclear region, will likely fall into the nuclear regions by the time of final coalescence~\citep{Souchay2012}. Therefore, the molecular gas in the central region will likely fall into the nuclear regions of NGC 6240 and contribute to the second starburst that will turn NGC 6240 into a bona fide ULIRG. However, the high velocity dispersion with respect to the vertical extent of the gas means it is also possible that the structure will dissipate prior to this streaming.  These observations and models shed light on one mechanism for star formation in major gas-rich mergers, that is, small-scale tidal bridges forming between progenitor galaxy nuclei that may ultimately feed into the nuclear regions.

\acknowledgments
This paper makes use of the following ALMA data: ADS/JAO.ALMA 2013.1.00813.S., 2015.1.00658.S. ALMA is a partnership of ESO (representing its member states), NSF (USA) and NINS (Japan), together with NRC (Canada), MOST and ASIAA (Taiwan), and KASI (Republic of Korea), in cooperation with the Republic of Chile. The Joint ALMA Observatory is operated by ESO, AUI/NRAO and NAOJ. The National Radio Astronomy Observatory is a facility of the National Science Foundation operated under cooperative agreement by Associated Universities, Inc.

\software{CASA (\cite{McMullin2007}), LIME (\cite{Brinch2010})}



\end{document}